\documentclass[10pt,letterpaper]{article}
\usepackage[top=0.85in,left=2.75in,footskip=0.75in,marginparwidth=2in]{geometry}

\usepackage[utf8]{inputenc}

\usepackage{mathpazo}
\usepackage{times}
\usepackage{amsmath}
\usepackage{amsfonts}
\usepackage{latexsym}
\usepackage{amssymb}
\usepackage{mathabx}
\usepackage{float}
\usepackage{upref}
\usepackage{theorem}
\usepackage{graphicx}
\usepackage{psfrag}
\usepackage{cite}
\usepackage{multirow}
\usepackage{enumerate}
\usepackage{enumitem}
\usepackage{algpseudocode}

\usepackage{mdframed}
\usepackage{mathtools}

\usepackage[hidelinks]{hyperref}
\usepackage{graphicx,subfig}
\usepackage[T1]{fontenc}
\usepackage{wrapfig}
\usepackage[font=footnotesize,labelfont=bf]{caption}
\usepackage{soul}
\usepackage{times}

\usepackage{color}
\usepackage{algorithm,algpseudocode}

\newcommand{\eps}{\epsilon}
\theoremstyle{plain}
\theorembodyfont{\normalfont\slshape}
\newtheorem{thm}{Theorem$\!$}

\newtheorem{clm}[thm]{Claim$\!$}

\newtheorem{lem}[thm]{Lemma$\!$}

\newtheorem{prop}[thm]{Proposition$\!$}

\newtheorem{cor}[thm]{Corollary$\!$}

\newtheorem{defn}[thm]{Definition$\!$}

\newtheorem{xmpl}[thm]{Example$\!$}

\newtheorem{cnstr}{Construction$\!$}

\newtheorem{rmk}[thm]{Remark$\!$}

\renewcommand{\leq}{\leqslant}

\renewcommand{\geq}{\geqslant}

\newcommand{\Cref}[1]{Co\-ro\-lla\-ry\,\ref{#1}}

\newcommand{\E}{{\Bbb E}}

\def\P{\mathbb{P}}
\def\E{\mathbb{E}}

\outer\def\proclaim #1. #2\par{\medbreak
	\noindent{\bf#1.\enspace}{\sl#2\par}%
	\ifdim\lastskip<\medskipamount \removelastskip\penalty55\medskip\fi}

\usepackage{cite}

\usepackage{nameref,hyperref}

\usepackage[right]{lineno}

\usepackage{microtype}
\DisableLigatures[f]{encoding = *, family = * }

\usepackage{changepage}

\usepackage[aboveskip=1pt,labelfont=bf,labelsep=period,singlelinecheck=off]{caption}

\makeatletter
\renewcommand{\@biblabel}[1]{\quad#1.}
\makeatother

\usepackage{lastpage,fancyhdr,graphicx}
\usepackage{epstopdf}
\fancyheadoffset[L]{2.25in}
\fancyfootoffset[L]{2.25in}

\usepackage{xcolor}

\usepackage{color}

\definecolor{Gray}{gray}{.25}

\usepackage{graphicx}

\usepackage{sidecap}

\usepackage{wrapfig}
\usepackage[pscoord]{eso-pic}
\usepackage[fulladjust]{marginnote}
\reversemarginpar

\begin{document}
\vspace*{0.35in}

\begin{flushleft}
{\Large
\textbf\newline{Semi-Quantitative Group Testing for Efficient and Accurate qPCR Screening of Pathogens with a Wide Range of Loads}
}
\newline
\\
Ananthan Nambiar\textsuperscript{1, $\dagger$},
Chao Pan\textsuperscript{2,3, $\dagger$},
Vishal Rana\textsuperscript{2, $\dagger$},
Mahdi Cheraghchi\textsuperscript{4},
Jo\~ao Ribeiro\textsuperscript{5},
Sergei Maslov\textsuperscript{1,3,*},
Olgica Milenkovic\textsuperscript{2,3,*}
\\
\bigskip
1. Department of Bioengineering, University of Illinois Urbana-Champaign, Urbana, Illinois, USA.
\\
2. Department of Electrical and Computer Engineering, University of Illinois Urbana-Champaign, Urbana, Illinois, USA.
\\
3. Center for Artificial Intelligence and Modeling, Carl R. Woese Institute for Genomic Biology, University of Illinois Urbana-Champaign, Urbana, Illinois, USA.
\\
4. Department of Electrical Engineering and Computer Science, University of Michigan, Ann Arbor, Michigan, USA.
\\
5. NOVA  LINCS and NOVA School of Science and Technology, Caparica, Portugal.

\bigskip
$\dagger$ These authors contributed equally and their names are listed in alphabetical order.\\
* milenkovic@illinois.edu, maslov@illinois.edu

\end{flushleft}

\section*{SUMMARY}
Pathogenic infections pose a significant threat to global health, affecting millions of people every year and presenting substantial challenges to healthcare systems worldwide. Efficient and timely testing plays a critical role in disease control and transmission prevention. Group testing is a well-established method for reducing the number of tests needed to screen large populations when the disease prevalence is low. However, it does not fully utilize the quantitative information provided by qPCR methods, nor is it able to accommodate a wide range of pathogen loads. 
To address these issues, we introduce a novel adaptive semi-quantitative group testing (SQGT) scheme to efficiently screen populations via two-stage qPCR testing. The SQGT method quantizes cycle threshold ($Ct$) values into multiple bins, leveraging the information from the first stage of screening to improve the detection sensitivity. Dynamic $Ct$ threshold adjustments mitigate dilution effects and enhance test accuracy. Comparisons with traditional binary outcome GT methods show that SQGT reduces the number of tests by $24$\% while maintaining a negligible false negative rate.


\section*{INTRODUCTION}

Pathogenic infections in humans can cause a wide range of diseases, from mild ailments like the common cold or strep throat to more severe and life-threatening illnesses such as COVID-19, Ebola, and Tuberculosis~\cite{alberts2002introduction, baker2022infectious}. These diseases are spread through the proliferation of pathogens within the host and subsequent transmission to other susceptible individuals, often leading to an outbreak in a population. The amount of pathogen in a host, typically referred to as the viral load in the case of viruses, is most frequently expressed in terms of the number of pathogen particles per milliliter of the collected fluid sample. It can vary significantly from the time of infection until recovery and can correlate with the severity of symptoms~\cite{fraser2007variation, zheng2020viral, fajnzylber2020sars}. To quantify viral loads, the real-time reverse transcription-polymerase chain reaction (qPCR) method is widely used, which reports the number of amplification cycles before the amount of genetic material in the sample reaches a prescribed threshold for detection, known as the cycle threshold or $Ct$ value.

Individual samples are usually tested using qPCR to monitor disease progression in patients, but when screening a population for infected individuals, it is more efficient to test large groups of samples simultaneously. Group testing (GT) is a strategy that involves pooling multiple samples prior to running qPCR tests, and subsequently detecting infected individuals in the groups based on the test results. This reduces the overall number of tests required while minimizing the false negative rate (FNR), which is critical in infectious disease screening methods, as undetected positive individuals can lead to the rapid spread of disease. Various GT strategies have been proposed in the past to increase the efficiency of wide-scale testing~\cite{dorfman1943detection, indyk2010efficiently, eberhardt2020multi}, which are implemented using adaptive or non-adaptive protocols. Adaptive testing allows for the sequential selection of groups, while non-adaptive testing requires the selection of all test groups at the same time.

The first known GT scheme, proposed by Dorfman~\cite{dorfman1943detection}, is an example of adaptive GT with binary outcomes (positive or negative), and is not designed to use the quantitative information about viral load. However, fully quantitative testing schemes, including compressive sensing~\cite{donoho2006compressed, ghosh2021compressed}, are susceptible to measurement noise, require specialized pooling matrices, and come with performance guarantees only when the ratio of maximum to minimum viral load is confined to a relatively narrow interval~\cite{aeron2010information}. This is not the case for many viruses, including SARS-CoV-2, where viral loads of patients may differ by multiple orders of magnitude~\cite{fajnzylber2020sars}. Furthermore, the pooling of samples in both GT and compressive sensing methods leads to dilution, which can adversely impact the accuracy of test outcomes and cannot be directly addressed in a compressive sensing setting.

To address these limitations, we propose a new adaptive semi-quantitative group testing (SQGT) scheme that uses $Ct$ values quantized into more than two bins in a structured way. In addition, our scheme combines test outcomes from two rounds to improve the likelihoods of subjects being labelled correctly. To handle the dilution effect, we define multiple $Ct$ thresholds and dynamically adjust them based on the group size. Since GT was used during the COVID-19 pandemic, multiple theoretical approaches mostly based on Dorfman's method have been developed~\cite{yelin2020evaluation, alcoba2021increasing}. At the same time, several large-scale GT data sets containing $Ct$ values in COVID-19 infected individuals have been generated and made publicly available \cite{barak2021lessons,de2020sample, hogan2020sample}. Therefore we test our SQGT scheme on COVID-19 data and compare it Dorfman's method, showing an increase in testing efficiency. For example, for a population infection rate of $0.02$, our SQGT method uses $24$\% fewer tests than the binary outcome Dorfman's GT method, while maintaining a negligible FNR compared to qPCR noise.

\section*{ALGORITHMS AND RESULTS} \label{sec:sqgt}
\subsection*{Basics of Group Testing}
\begin{figure}[h]
    \centering
    \includegraphics[width=0.5\linewidth]{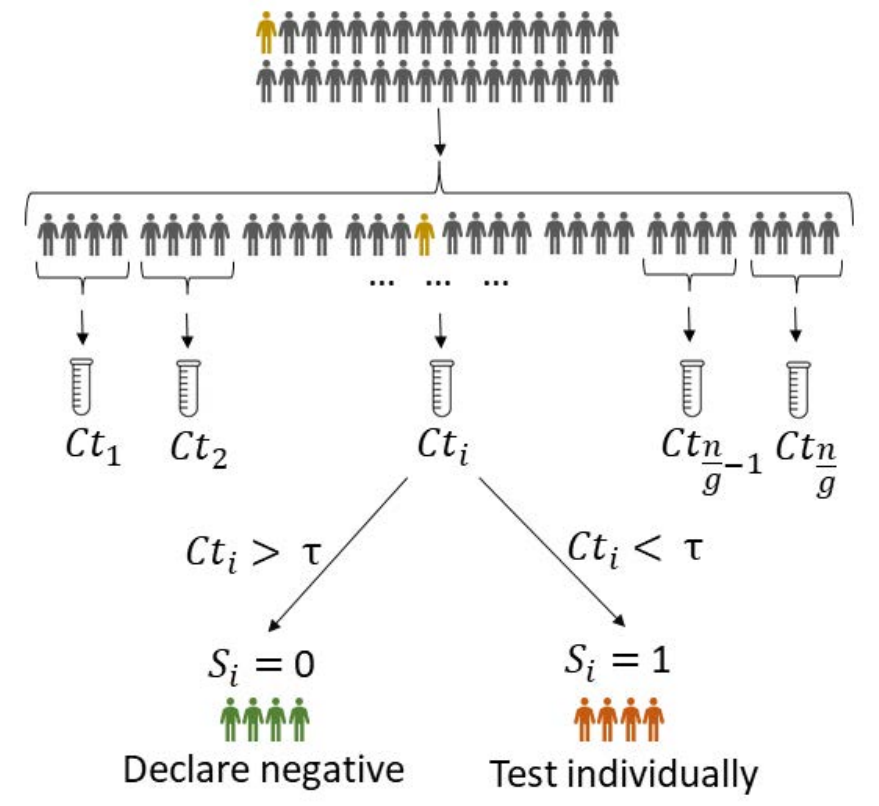}
    \caption{Dorfman's two-stage GT protocol. The test subjects are randomly partitioned into groups of optimized size $g$ and tested as a group. All individuals in positive groups are subsequently tested individually. As before, 
 $Ct$ stands for the cycle threshold value of the group under consideration. Note that this GT protocol only uses a binary decision variable, yes ($1$) and no ($0$), for the case that $Ct<\tau$ and $Ct>\tau$, respectively. The decision threshold $\tau$ depends on the protocol used for qPCR.}
    \label{fig:dorfman_basic}
\end{figure}

Group testing (GT), in its most basic form, performs screening of a collection of potentially positive individuals by splitting them into test groups involving more than one individual so as to save on the total number of tests performed. The outcome of a group of test subjects is interpreted as follows: the result is declared positive (and denoted by $1$) if at least one of the individuals in the tested group is infected; and, the test result is declared negative (and denoted by $0$) if there are no infected individuals in the group. From a theoretical point of view, GT aims to find an optimal strategy for grouping individuals so that the number of binary tests needed to accurately identify all infected individuals is minimized. GT can be implemented using nonadaptive and adaptive approaches. Unlike adaptive GT, nonadaptive schemes require that all tests are performed simultaneously so that the outcome of one test cannot be used to inform the selection of individuals for another test. The first known GT scheme by Dorfman~\cite{dorfman1943detection} is an example of adaptive screening since it involves two stages of testing, one of which isolates groups with infected individuals, and another one that identifies the actual infected individuals. In general, adaptive schemes use multiple stages of testing and different combinations of individuals to best inform the sequence of tests to be made. When specializing Dorfman's scheme for qPCR screening, the decision about positive and negative group labels is made based on $Ct$ values (see Figure~\ref{fig:dorfman_basic}).

Despite their widespread use, GT methods have notable shortcomings when used in systems that provide more quantitative information than a binary answer of the form ``yes-no,'' such as is the case for qPCR screening. This motivates developing extensions of GT schemes that make use of the more quantitative information available from experiments. When all of the available quantitative information is used, the generalized GT scheme represents a form of compressive sensing (CS)~\cite{donoho2006compressed, dai2009subspace, candes2006stable}. However, CS-based schemes require the ratio of the maximum and minimum pathogen concentrations to be properly bounded~\cite{aeron2010information}. This type of assumption does not hold for a large number of infectious diseases, including COVID-19, where the viral concentrations can vary over several orders of magnitude~\cite{fajnzylber2020sars}. In the presence of infected individuals with widely different loads, CS approaches will mask individuals with low pathogen concentrations. 

\begin{figure}[h]
    \centering
    \includegraphics[width=\linewidth]{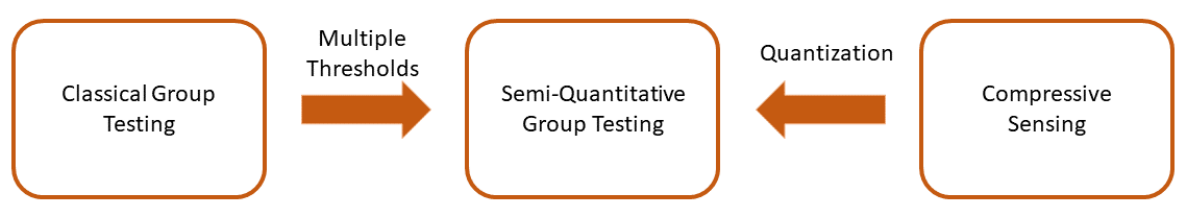}
    \caption{Semi-quantitative GT generalizes Dorfman's GT by using more than one threshold and, like CS, uses information about the estimate of the total number of infected individuals, but with the numbers quantized according to predetermined cluster selections.}
    \label{fig:gt_sqgt_cs}
\end{figure}

Here we propose a more structured approach to GT that straddles the classical Dorfman's scheme and fully quantitative CS approaches. Our semi-quantitative GT scheme (SQGT) can be seen as a multi-threshold version of Dorfman's GT with two independently permuted groups of samples or a quantized version of adaptive CS (see Figure~\ref{fig:gt_sqgt_cs}). More details are provided in the following subsection.

\subsection*{Semi-Quantitative Group Testing}

\begin{figure}[h]
    \centering
    \includegraphics[width=\linewidth]{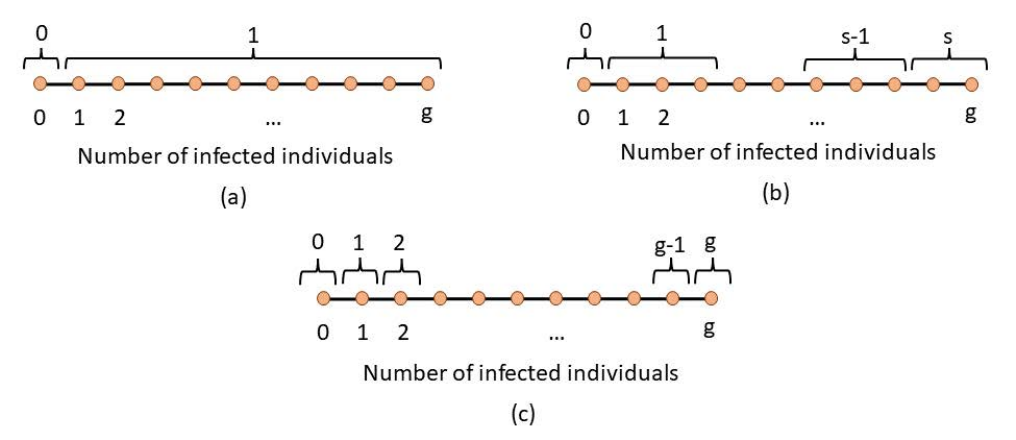}
    \caption{GT interpreted through quantitative output quantization. The quantitative output corresponds to the actual number of infected individuals in a group. In (a), corresponding to Dorfman's GT, the quantizer maps all outcomes involving more than one infected individual to a score $1$. The score $0$ indicates that there are no infected individuals in the group. In (b), corresponding to a general SQGT scheme, the quantizer is allowed to map any collection of outcomes to any choices of scores. This implies that the number of possible test results may be larger than two, but upper bounded by the size of the group $g$. The simplest version of SQGT based on a uniform quantizer is depicted in (c).}
    \label{fig:bin_gt}
\end{figure}

SQGT is a GT protocol that interprets test results as estimates of the number of infected individuals in each tested group. Broadly speaking, unlike Dorfman's GT which generates binary responses ($0$, for a noninfected group, and $1$ when at least one infected subject is present in the group, see Figure~\ref{fig:bin_gt} a), SQGT produces answers of the form ``between $x$ and $y$ infected individuals in the group'' (see Figure~\ref{fig:bin_gt} b). 
For qPCR experiments, the range of values for the number of infected individuals in the group may be estimated from the $Ct$ value of the group.

For a general SQGT scheme, one seeks a collection of $\geq 1$ measurement thresholds, such that the outcome of each test is an interval for the possible number of infected individuals, i.e., the outcome of an SQGT experiment specifies lower and upper bounds on the number of infected individuals in a group. If the thresholds are consecutive integers covering all possible options for the number of infected individuals in a group, the scheme reduces to additive (quantitative) GT~\cite{lindstrom1975determining,wolf1985born} (see Figure~\ref{fig:bin_gt} c). 

\begin{figure}[h]
    \centering
    \includegraphics[width=\linewidth]{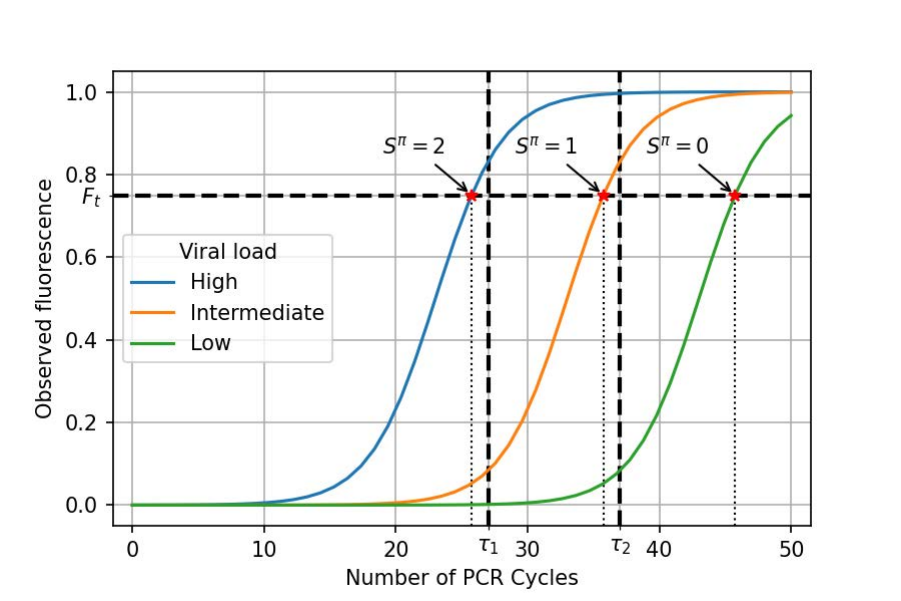}
    \caption{An example of qPCR amplification curves and two-threshold ($\tau_{1}, \tau_{2}$) SQGT. The two thresholds apply to $Ct$ values while the actual measurement corresponds to the intersection of the $F_t$ line (the fluorescence threshold) and the amplification curve. For example, the left-most red star indicates the intersection of the high viral load amplification curve with $F_t$ and the corresponding measurement falls into the quantization bin denoted by $S^{\pi} = 2$.}
    \label{fig:pcr_tresh}
\end{figure}

Although nonadaptive SQGT has been previously analyzed from an information-theoretic perspective~\cite{emad2014semiquantitative,emad2016code,cheraghchi2021semiquantitative}, practical implementations for adaptive SQGT schemes are still lacking, especially in the context of qPCR testing. Our approach is the first adaptive SQGT scheme that is specifically designed for real-world qPCR testing. It operates directly on the $Ct$ values and makes use of two  thresholds, $\tau_1$ and $\tau_2$ (see Figure~\ref{fig:pcr_tresh}). This choice for the number of thresholds balances the ease of implementation of a testing scheme in a laboratory with the ability to use the quantitative information from a qPCR test more efficiently\footnote{We also observe in practice that using more than two thresholds leads to diminishing returns in the number of tests saved but significantly increasing the complexity of the scheme.}. 

The main idea behind our $Ct$ value-based SQGT approach is to perform a two-stage SQGT protocol with randomly permuted groups of subjects and risk assessment based on the $Ct$ values obtained after the first stage. More specifically, the scheme involves the following three steps:

\begin{itemize}
\item First, we create two separate, randomly permuted lists of $n$ subjects. Each of these lists is then evenly divided into groups of a specified size, $g$, which are subsequently tested. It's important to underline that the ideal test group size, $g$, for our methodology may differ from that typically utilized in Dorfman's GT approach.
\item Second, since GT inevitably leads to sample dilution, we adjust the $Ct$ thresholds in the SQGT scheme to account for this effect. Note that each individual's sample contributes to two $Ct$ values: one from the group they were initially part of in the first permuted list, and another from their group in the second permuted list. This dual-measurement system provides a way for cross-linking the results.

\item Third, we examine the pair of $Ct$ values associated with the individuals to stratify them into low-risk, medium-risk, and high-risk categories. Based on the risk category, the individuals are either immediately declared negative, or tested once again individually. Although the number of tests performed can be reduced by performing nonadaptive SQGT testing on all risky subjects (discussed in the Supplement Section~\ref{supp:nonadaptiveSQGT}), for simplicity we opt for individual testing.
\end{itemize}

Next, we describe our scheme in detail. We consider a population of $n$ individuals, arranged into groups of size $g$, and denote the fraction of infected individuals by $p$. Again, we only make use of two quantization thresholds, denoted by $\tau_1$ and $\tau_2$. Our scheme consists of two stages.

In the first stage, we group the patient samples into groups of size $g$, ensuring that each individual contributes to two different groups. To achieve this, we use two random permutations, $\pi_1$ and $\pi_2$, of the $n$ individuals so that they appear in different random orders. Subsequently, the ordered lists are split into groups of $g$ consecutive samples (for simplicity, we assume that $n$ is a multiple of $g$). The resulting groups are denoted by $\gamma^{\pi_1}_1, \gamma^{\pi_1}_2, \dots, \gamma^{\pi_1}_{n/g}$ and $\gamma^{\pi_2}_1, \gamma^{\pi_2}_2, \dots, \gamma^{\pi_2}_{n/g}$. It is important to note that each individual belongs to two groups, $\gamma^{\pi_1}_{i}$ and $\gamma^{\pi_2}_{j}$ with $i \in \{{1,\ldots,n/g\}}$ and $j \in \{{1,\ldots,n/g\}}$, where the two groups are created based on the two permuted lists. For both collections of groups, we perform separate qPCR experiments, denoting the outcomes as $Ct^{\pi_1}_{i}$ and $Ct^{\pi_2}_{j}$, respectively. Then we quantize the $Ct$ values into bins, and assign the test scores $S^{\pi_1}_i$ for group $\gamma^{\pi_1}_{i}$ and $S^{\pi_2}_j$ for group $\gamma^{\pi_2}_{j}$ using the threshold rule:

\begin{equation}
S^{\pi} = 
\begin{cases}
0, & \text{ if } Ct^{\pi} \geq \tau_2;\\
1, & \text{ if } \tau_1 < Ct^{\pi} < \tau_2;\\
2, & \text{ if } Ct^{\pi} \leq \tau_1.
\end{cases} \label{eq:thresh}
\end{equation}

Consequently, each individual is labeled by a pair of test scores $(S^{\pi_1}_i, S^{\pi_2}_j)$, representing the outcomes of the two group tests (for group $\gamma^{\pi_1}_{i}$ and $\gamma^{\pi_2}_{j}$) that the individual is involved in. We omit the subscripts $i$ and $j$ in the later context for simplicity of notation.

\begin{figure}[h]
    \centering
    \includegraphics[width=\linewidth]{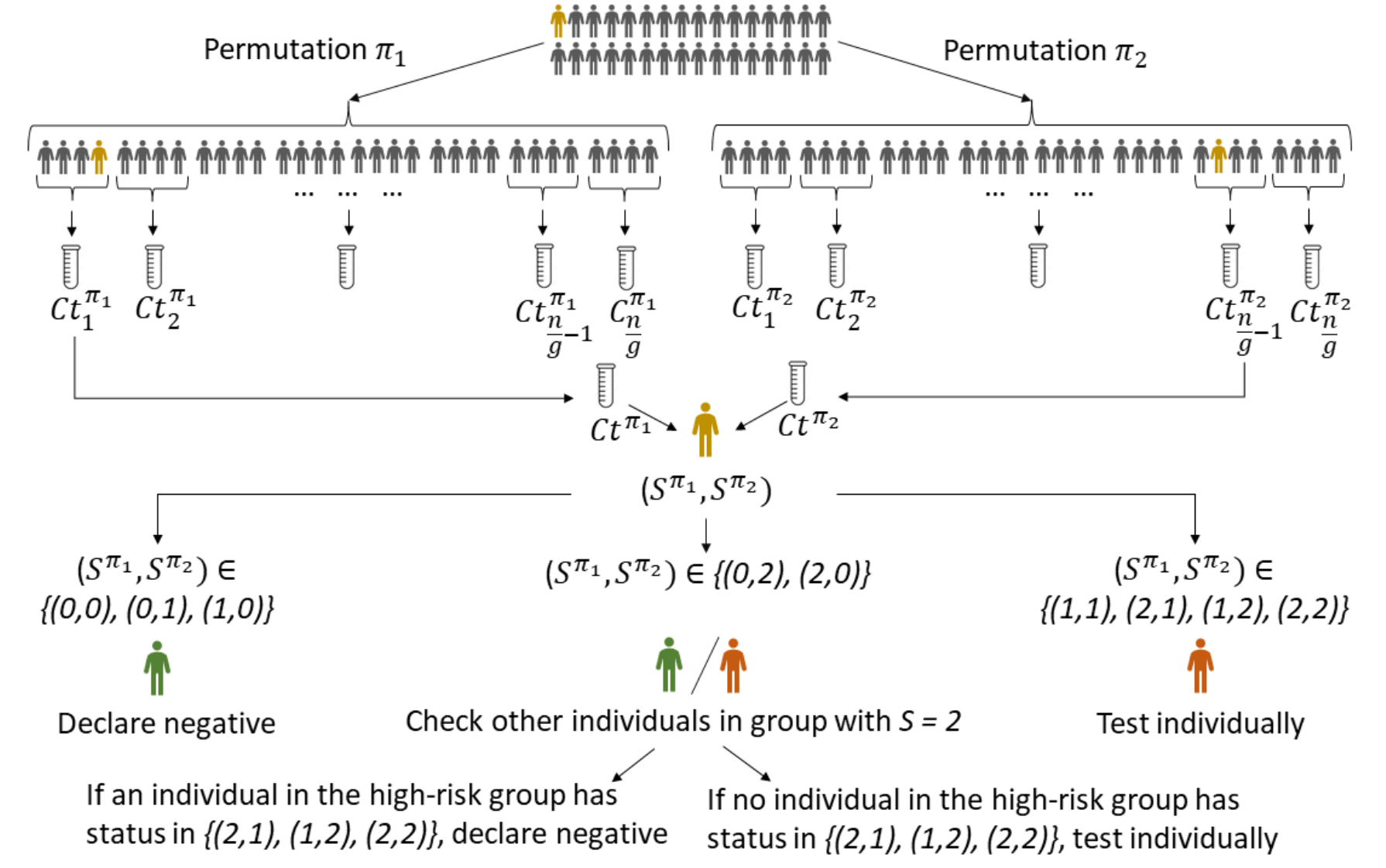}
    \caption{Our proposed two-stage SQGT scheme with two thresholds, as described in Equation~\ref{eq:thresh}. The approach is to run two parallel rounds of Dorfman-like group tests. To assess if the individual marked in orange is infected, we test them in two different groups, and collect the scores $(S^{\pi_1}, S^{\pi_2})$. Based on this pair of scores, we decide if the individual marked in orange needs to be individually tested or not. See the text for more details.}
    \label{fig:sqgt_basic}
\end{figure}

In the second stage, we classify individuals based on their scores $(S^{\pi_1}, S^{\pi_2})$. Individuals with scores $\{(0,0), (0,1), (1,0)\}$ are deemed low-risk and declared negative. In particular, scores $\{(0,1), (1,0)\}$ are declared to correspond to negative subjects because they were involved in a negative test group (score $0$) and intermediate $Ct$ value group (score $1$). Subjects with scores $\{(1,1), (2,1), (1,2), (2,2)\}$ are classified as high-risk and tested individually in a second stage of tests. For the remaining score pairs, $\{(2,0), (0,2)\}$, we proceed as follows: If the group with score $2$ contains another individual with a score in $\{(1,2),(2,1),(2,2)\}$, we classify the first individual as negative; otherwise, we conduct an individual test. We choose this option since it is unlikely that the first individual was positive, given the existence of even worse-scoring individuals in the same group. Figure~\ref{fig:sqgt_basic} illustrates the proposed two-stage SQGT scheme, while Figure~\ref{fig:dorfman_basic} depicts Dorfman's GT scheme. Supplement Sections~\ref{supp:modelGT} and ~\ref{supp:variableload} provide a detailed mathematical analysis of the various GT schemes discussed.

It is worth noting that conducting individual testing, as in the second stage of our SQGT scheme for the high-risk group, is suboptimal from the point of minimizing the number of tests. This issue is not limiting the application of the scheme since one can use a nonadaptive GT scheme in the second stage, thereby significantly reducing the number of second-stage tests. Since nonadaptive GT is conceptually more involved and harder to implement in practice than the above procedure, pertinent explanations are delegated to Supplement Section~\ref{supp:nonadaptiveSQGT}.

As we will demonstrate in the Results section, the proposed two-stage SQGT approach offers substantial reductions in the number of tests when compared to Dorfman-type tests. It remains to see if the reduction in the number of tests leads to undesirable increases in the FNR of the scheme. To address this question, we need to consider the influence of dilution effects on the test results and how one could adjust quantization thresholds to counter these effects.

\subsection*{Dilution Effects}
In most experiments involving GT, the test samples come in specified unit concentrations that are equal across all test subjects. This means that a group sample involving $g$ individuals will only use a fraction $1/g$ of the unit sample from each individual. This inevitably leads to dilution of the group sample, the level of which depends on the number of infected individuals in that particular group. When there is only a small number of infected individuals in the group, the overall viral load of the group sample may be lower than the detection threshold, thereby leading to highly undesirable false negatives. False negative rate (FNR) is related to true positive rate (TPR) through $\text{FNR}=1-\text{TPR}$, and the TPR function is often referred to as the sensitivity function.

A mathematical model for dilution effects was first proposed in~\cite{hwang1976group}, which introduced a special TPR function $TPR(p, g, d)$ of the form
\begin{linenomath}
\begin{align}\label{eq:hwang_model}
TPR(p, g, d) &= \mathbb{P}(\text{test result is declared positive}|\text{there is at least $1$ positive subject in the group}) \notag\\
&= p\left[1-(1-p)^{g^d}\right]^{-1}.
\end{align}
\end{linenomath}
Here, $p$ denotes the infection rate, $g$ denotes the group size, and $d$ denotes a parameter capturing the dilution level. When $d=0, TPR(p,g,0)=1$, indicating that there is no dilution; when $d=1$ and $g$ is large, $TPR(p,g,1)\sim p$, indicating that the sample is fully diluted and that the probability of correctly identifying a defective group is the same as the infection rate. More details on the TPR model for SQGT can be found in Supplement Section~\ref{supp:hwangFPR}. 


Although the dilution model~(\ref{eq:hwang_model}) is mathematically elegant and tractable for analysis, it provides a poor match for real-world measurements (see Figure~\ref{fig:fnr} (b)). A more practical approach to quantifying dilution effects is to assess how dilution impacts the actual viral load in a group. The empirical studies~\cite{jones2021estimating,barak2021lessons,brault2021group,de2020sample} consistently point out that the $Ct$ values of groups tend to be higher than the $Ct$ value of individual tests with high probability. This phenomenon is also due to dilution effects. Nevertheless, none of these works describe how to readjust the $Ct$ value used for declaring positives in the presence of dilution. In the context of SQGT, this is an even more important issue as the increased $Ct$ values can lead to degradation in the detection rate as well as a significantly increased number of measurements. This motivates exploring the relationship between the value of the $Ct$ threshold used for an individual test and that used for a group test. For the worst-case scenario when there is only one infected individual in a group of size $g$, the group $Ct$ value takes the form
\begin{linenomath}
\begin{align}\label{eq:ct_viral_load}
    Ct &= -M \log_{10}({v}/{g}) + B\notag\\
            &= -M\log_{10}(v) + B + M\log_{10}(g),
\end{align}  
\end{linenomath}

where $v$ denotes the viral load of the infected individual, and $M$ and $B$ are positive values denoting the slope and the intercept for the PCR calibration curve~\cite{jones2021estimating}. The exact values of $M$ and $B$ need to be estimated from the experimental data. Equation~(\ref{eq:ct_viral_load}) characterizes the relationship between the viral load and the $Ct$ value, and it implies that compared to individual testing, the group $Ct$ value will be higher by $M\log_{10}(g)$. The implication of this observation is that for GT, we need to increase the $Ct$ thresholds by $M\log_{10}(g)$.

\subsection*{Controlling and Modelling FNRs of PCR Tests}
\begin{figure}[h]
    \centering
    \includegraphics[width=0.48\textwidth]{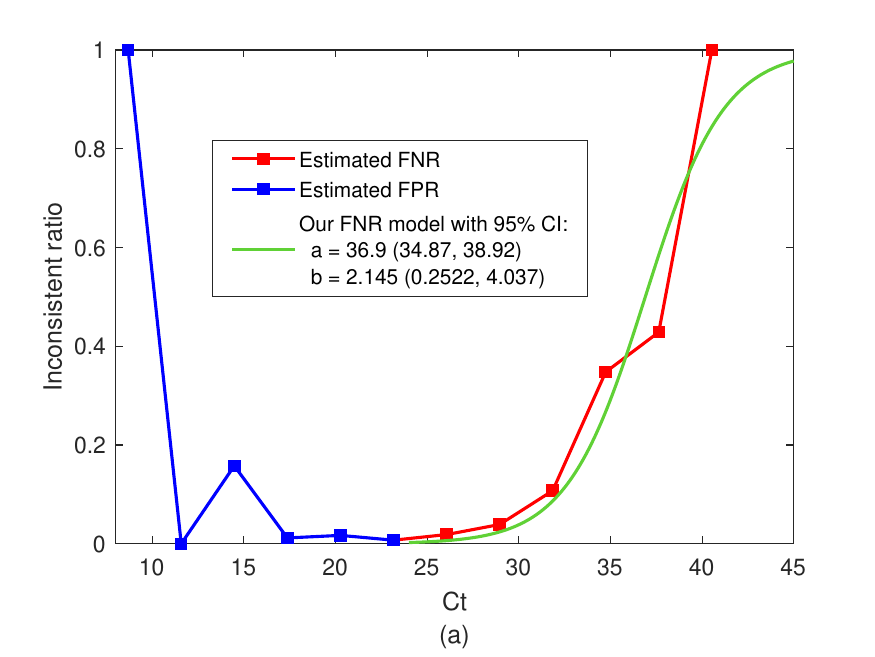}
    \includegraphics[width=0.48\textwidth]{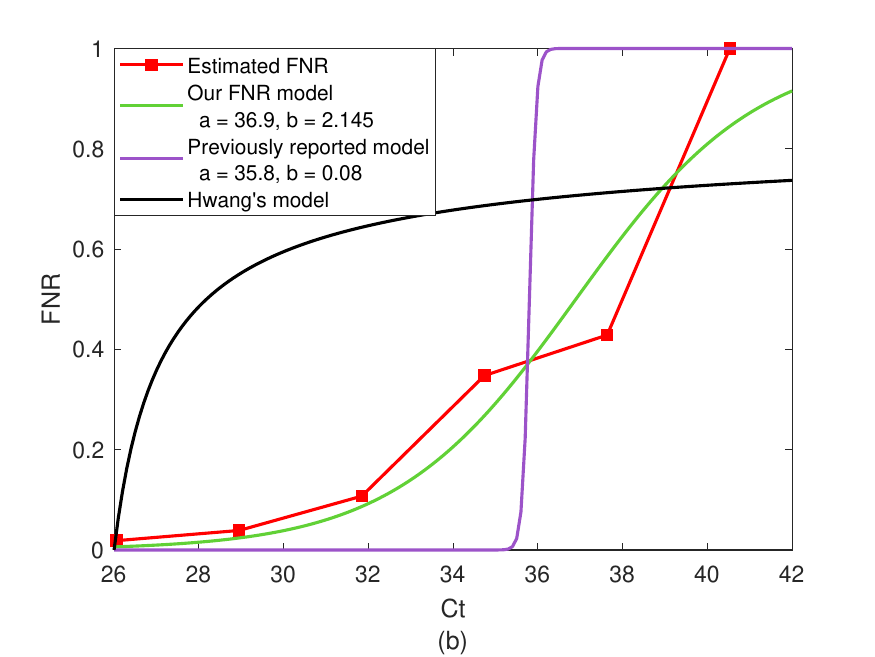}
    \caption{FNR estimated from data reported in~\cite{barak2021lessons} and different FNR models fitted to the real-world experimental data. (a) We count the cases where the group test was positive but all subjects individually tested negative. The ratio of the number of these ``inconsistent'' tests and the total number of tests with the same $Ct$ value is denoted as the ``inconsistent ratio''. Specifically, we consider the right half of the curve ($Ct>25$) to be caused by the false negative results, which agrees with the intuition that the FNR increases as the $Ct$ value increases. (b) We fit the FNR model from Equation~(\ref{eq:fnr}), and the ones from~\cite{hwang1976group,lin2020group} to the real-world experimental data. As it is apparent, the black and purple lines provide a poor fit to the data while our model (green line) with parameters $(a=36.9,b=2.145)$ represents a significantly more accurate fit.}
    \label{fig:fnr}
\end{figure}

In order to quantify the trade-off between the FNR and the reduction in the number of group tests when using the proposed SQGT scheme, we express the FNR, an important metric with respect to test accuracy, as a function of the $Ct$ value. For this purpose, we use the large-scale real-world GT dataset~\cite{barak2021lessons}. Our FNR model is based on the following ``sigmoid'' function,
\begin{linenomath}
\begin{align}\label{eq:fnr}
FNR(Ct)=\left[1+\exp\left(\frac{a-Ct}{b}\right)\right]^{-1},
\end{align}
\end{linenomath}
where $a,b$ are two tunable parameters that can be used to fit the measured/estimated FNRs. Note that similar ideas were also discussed in~\cite{lin2020group}; however, as may be seen from Figure~\ref{fig:fnr} (b), the FNR function ($a=35.8, b=0.08$) proposed in~\cite{lin2020group} significantly deviates from real-world experimental data.

In practice, the values of FNRs are hard to estimate as this requires multiple tests of the same individual. In the GT context, there are two ways to estimate FNRs. The direct scenario is to compute FNRs by counting the instances when a group test was negative but at least one member from that group tested positive. However, in all experimental verification of GT protocols, individuals whose group tested negative are eliminated from future retesting. This renders the direct approach impossible to pursue in practice. The indirect approach is to count the cases where the group test was positive but all subjects individually tested negative. In this work, we follow the second approach to estimate the FNRs. The ratio of the number of these ``inconsistent'' tests and the total number of tests with the same $Ct$ value is shown in Figure~\ref{fig:fnr} (a). Note that these results can correspond to either a false positive for the group test, or a false negative for one or more of the individual tests. Here we consider the right half of the curve ($Ct>25$) to be caused by the false negative results, which agrees with the intuition that the FNR increases as the $Ct$ value increases. Our fitted FNR curve is shown in Figure~\ref{fig:fnr} (b), along with the estimated FNR curve from experimental results, and the models from~\cite{hwang1976group,lin2020group}. As it is apparent, the latter provides a poor fit to the data while our model with parameters $(a=36.9,b=2.145)$ represents a significantly more accurate fit. 

The FNR shown in Figure~\ref{fig:fnr} corresponds to individual tests, for which we do not know the correct $Ct$ values. Therefore, we shift the group test $Ct$ values by $M\log_{10}(g)=2.895$ in Equation~(\ref{eq:ct_viral_load}) to estimate the individual $Ct$ values. A detailed discussion of the data processing and FNR estimation pipeline is included in the Methods Section.

\subsection*{Case Study of the SQGT Protocol Applied to COVID-19 Data}
While the SQGT framework is broadly applicable to PCR-based pathogen screening, general data is usually limited for pathogens other than SARS-CoV-2. The COVID-19 pandemic has resulted in an unprecedented amount of publicly available qPCR test data, which motivates testing our SQGT framework on real-world SARS-CoV-2 data. Our reported results pertain to a set of $133,816$ SARS-CoV-2 $Ct$ values of qPCR tests performed in Israel between April and September 2020 as reported in~\cite{barak2021lessons}. To explore a range of different infection scenarios without performing additional experiments, we simulated populations of $10,000$ individuals of which a fraction $p \in \{0.02, 0.05, 0.1\}$ was infected by the virus. The $Ct$ values of the infected individuals were randomly sampled from the real-world dataset of~\cite{barak2021lessons}, and converted into estimated viral loads using Equation~\ref{eq:lineari} (see also the Methods section). The viral loads of uninfected individuals were set to $0$. 

Following the SQGT scheme of Figure~\ref{fig:sqgt_basic}, samples are partitioned into groups of $g$ individuals whose viral loads were subsequently averaged and converted to $Ct$ values as described in the Methods section (Equation~\ref{eq:linear}). Following standard diagnostic procedures, individuals were declared negative if their $Ct$ values exceeded a threshold (in our case, set to $37$ as suggested in \cite{hu2020factors}).


To analyze the magnitude of the savings in the number of tests required for the GT scheme compared to individual screening, independent of PCR assay noise, we ran both Dorfman's GT and SQGT on the model data. The tests were performed under the assumption that qPCR assays are error-free. Supplement Figure~\ref{fig:error-free} shows these results for all three infection rates $p$. We performed a sweep of group sizes $g$ for each value of $p$ to identify their optimal values. While both GT schemes require significantly fewer than the $10,000$ tests needed for individual testing, SQGT consistently outperforms Dorfman's GT for all three infection rate levels. In addition, Supplement Figure~\ref{fig:error-free} shows that the group-dependent thresholds help to avoid false negatives that would have occurred due to dilution effects, as expected.

\begin{figure}[ht!]
     \centering
     \includegraphics[width=1\textwidth]{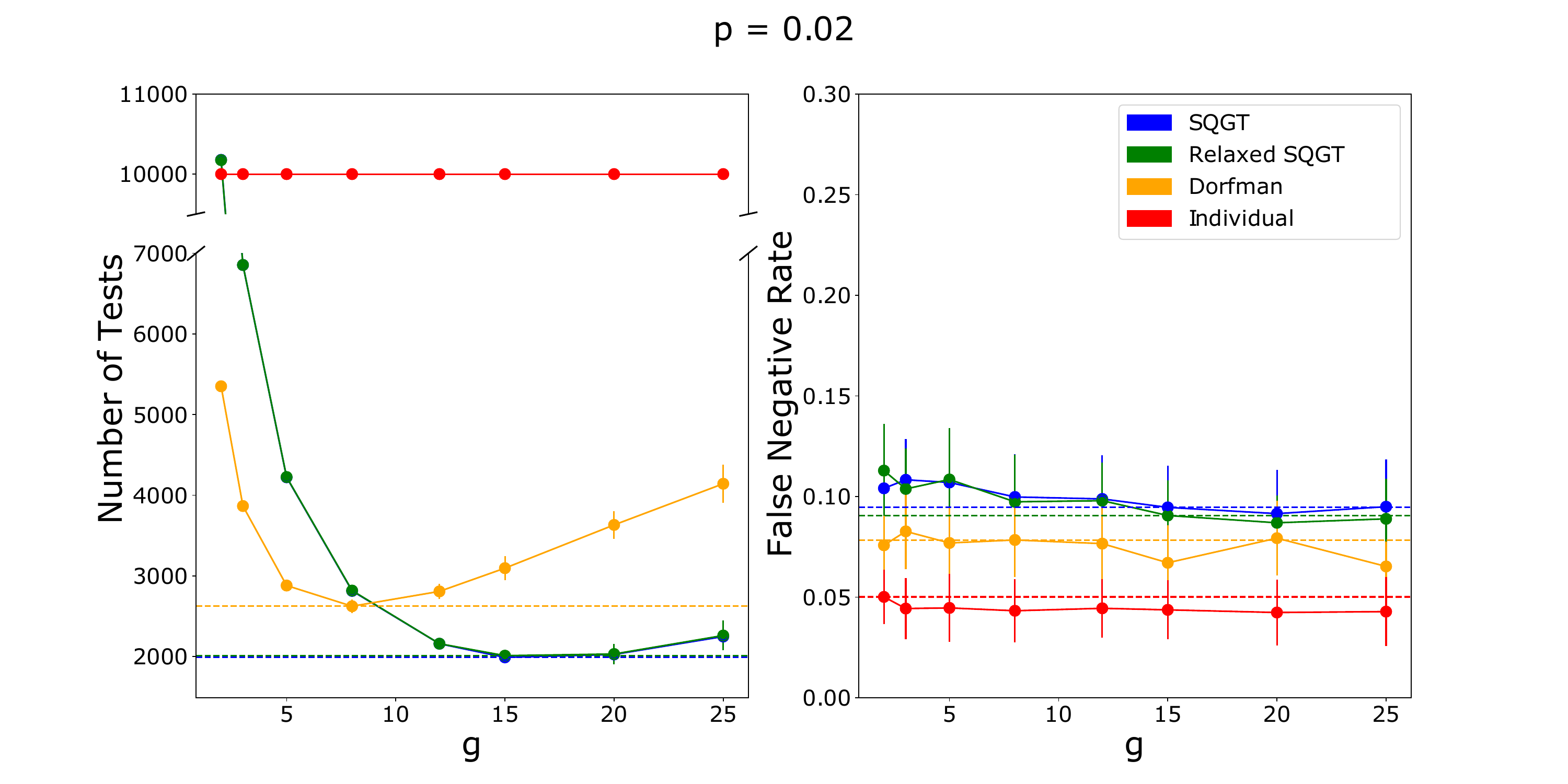}\\
     \includegraphics[width=1\textwidth]{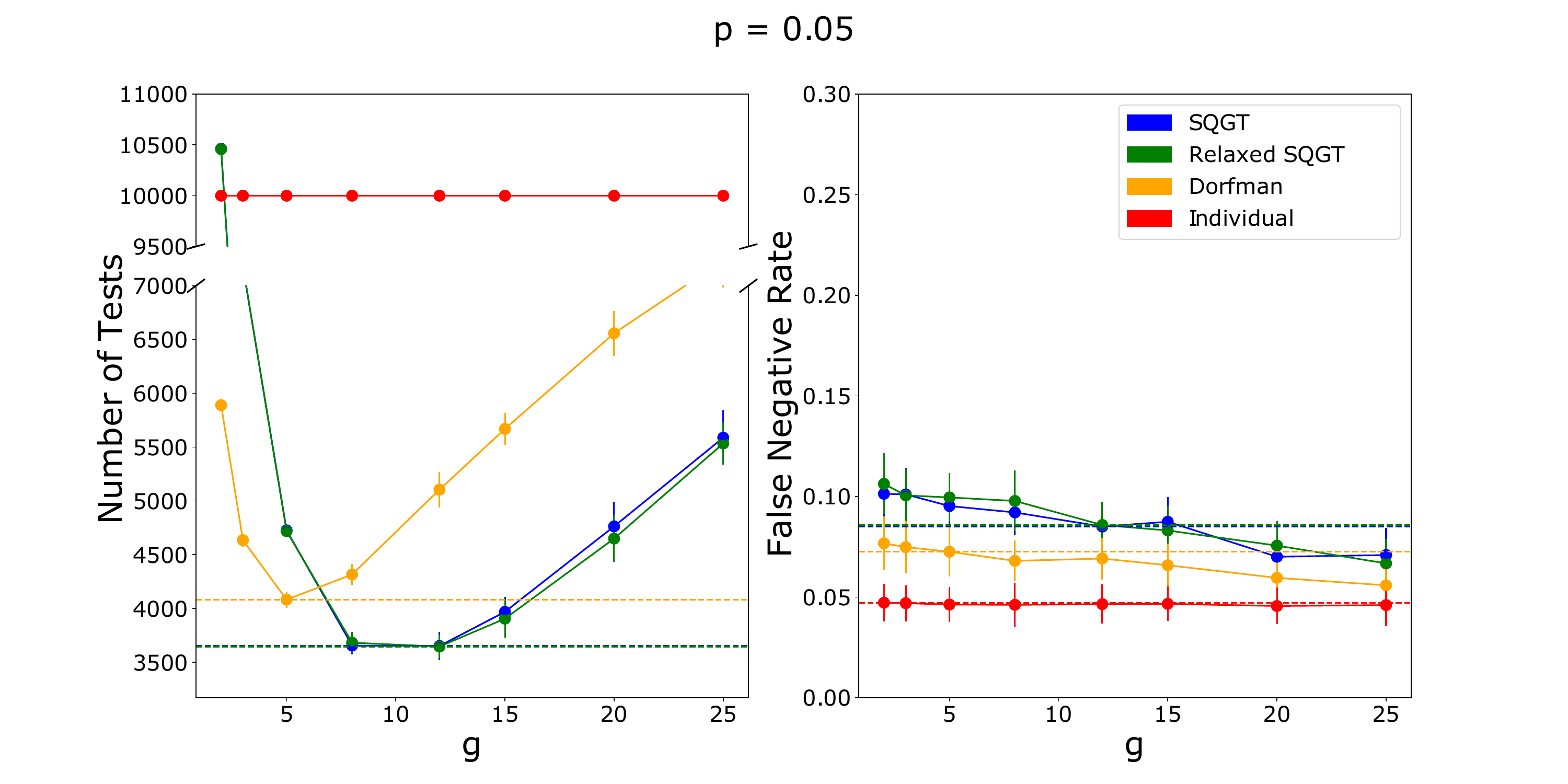}\\
     \includegraphics[width=1\textwidth]{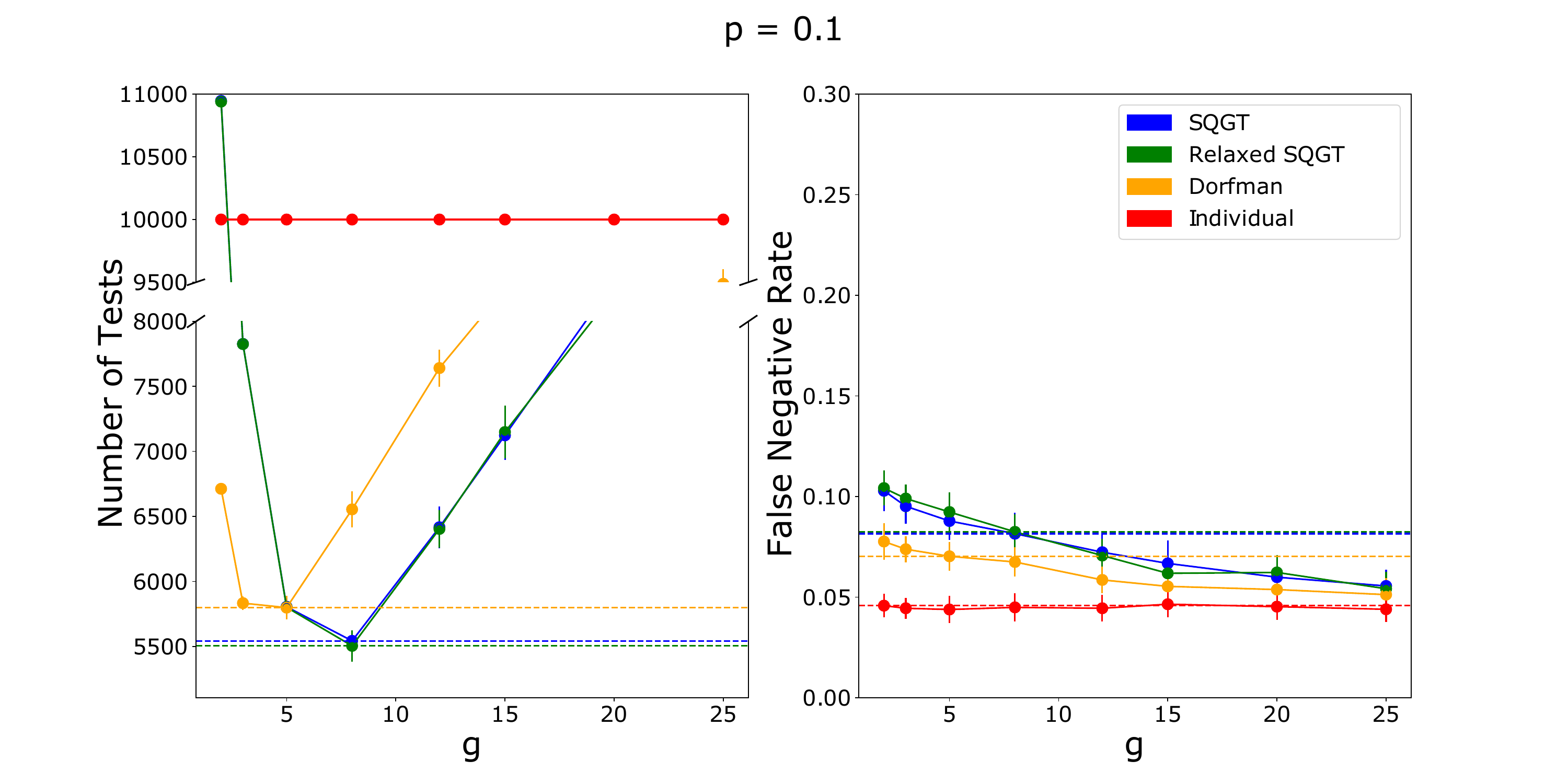}
     \caption{The number of tests used and the FNRs of the SQGT protocol (blue), Dorfman's GT (orange), and individual testing (red) for infection rates $p\in \{0.02,0.05,0.1\}.$ The dashed lines show the number of tests and FNRs for the optimal group size (i.e., the group size that minimizes the number of tests needed) for each scheme. }
     \label{fig:noisy}
 \end{figure}

However, as noted in the previous section, qPCR assays are not error-free in practice, and as a result, the false negatives in GT schemes could be due to either dilution effects or qPCR noise. Therefore, we incorporated qPCR noise into our model to make it more realistic. This was done by including the empirically fitted FNR in Figure~\ref{fig:fnr} into the PCR assays in our model (see the Methods section for details). Figure~\ref{fig:noisy} shows that while the noise has very limited effects on the number of tests required by each GT scheme, it does have the expected effect of increasing the FNR of both individual and group tests. For individual testing, the noise function we fit appears to correspond to an FNR of just under $0.05$, which is comparable to the empirically determined values reported in~\cite{TAKAHASHI2022e01496} and~\cite{woloshin2020false}. The FNR values of both GT schemes are also consistently slightly higher than those of individual testing. To compare the FNR of SQGT and Dorfman's GT, we first identify the optimal group size for each scheme by picking the value $g$ for which the scheme requires the least number of tests. When $p=0.02$, the optimal value of $g$ for SQGT was $15$ with an average of $1,989.8$ tests required; at the same time, Dorfman's GT required $2,623.6$ tests for an optimal group size $g=8$. These optimal group sizes correspond to FNRs of about $0.0946$ for SQGT and $0.0784$ for Dorfman's GT, respectively. When the infection rate is increased to $0.05$, the optimal group sizes are smaller, with $g=12$ and $g=5$ for SQGT and Dorfman's GT, respectively. These group sizes correspond to $3,651.7$ tests with an FNR of $0.851$ for SQGT and $4,082.6$ tests with an FNR of $0.726$ for Dorfman's GT. Finally, at $p=0.1$ the optimal group size for SQGT was identified as $g=8$, with $5,542.2$ tests and an FNR of $0.815$, while for Dorfman's GT the results indicated $g=5$, with $5,798.0$ tests and an FNR of $0.703$. The observed trend is that SQGT offers savings in the number of tests at the expense of a slight increase in FNR. It should also be noted that this increase is often within the error-bounds of the FNRs.

\begin{figure}
    \centering
    \includegraphics[width=0.8\textwidth]{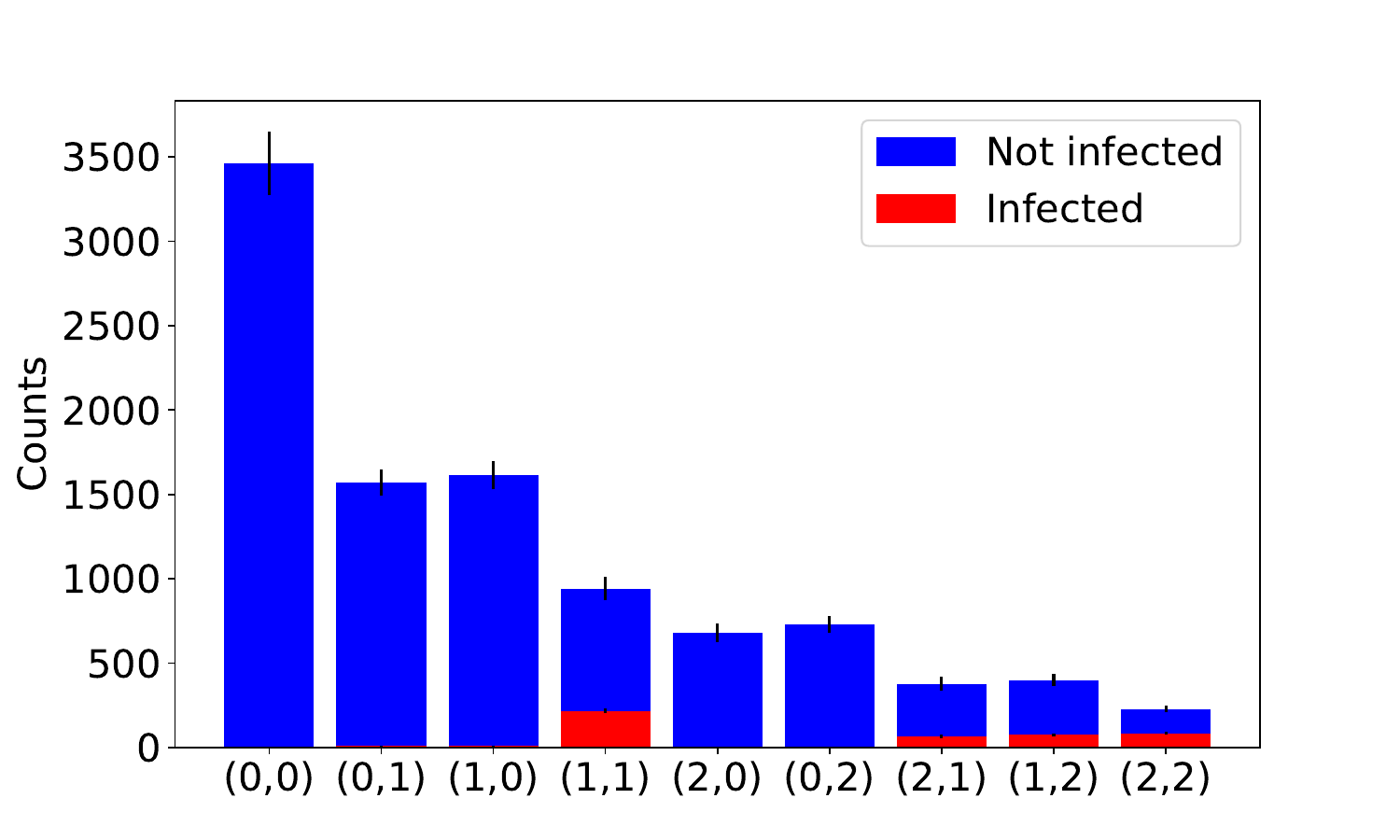}
    \caption{The number of individuals with each possible outcome for the pair of test results in the SQGT scheme. The number of infected individuals is shown in red, while the number of healthy (uninfected) individuals is shown in blue.}
    \label{fig:breakdown}
\end{figure}

In addition, we tested a modified version of SQGT where individuals with a $(2,0)$ or $(0,2)$ result are declared negative without further testing. As shown in Figure~\ref{fig:noisy}, this version of the SQGT method performs similarly to the regular SQGT. To investigate the reason behind this finding, we plotted the number of individuals for each possible outcome of the SQGT scheme for an infection rate of $0.04$ and the corresponding optimal group size $g=12$. As can be seen in Figure~\ref{fig:breakdown}, the $(2,0)$ and $(0,2)$ test results consist only of uninfected individuals. Therefore, it makes sense that declaring them negative without further testing has no effect on the FNR. For a mathematical analysis of the phenomena and related GT models, the reader is referred to Supplement Section~\ref{supp:modelGT}. 

\begin{figure}
    \centering
    \includegraphics[width=1\textwidth]{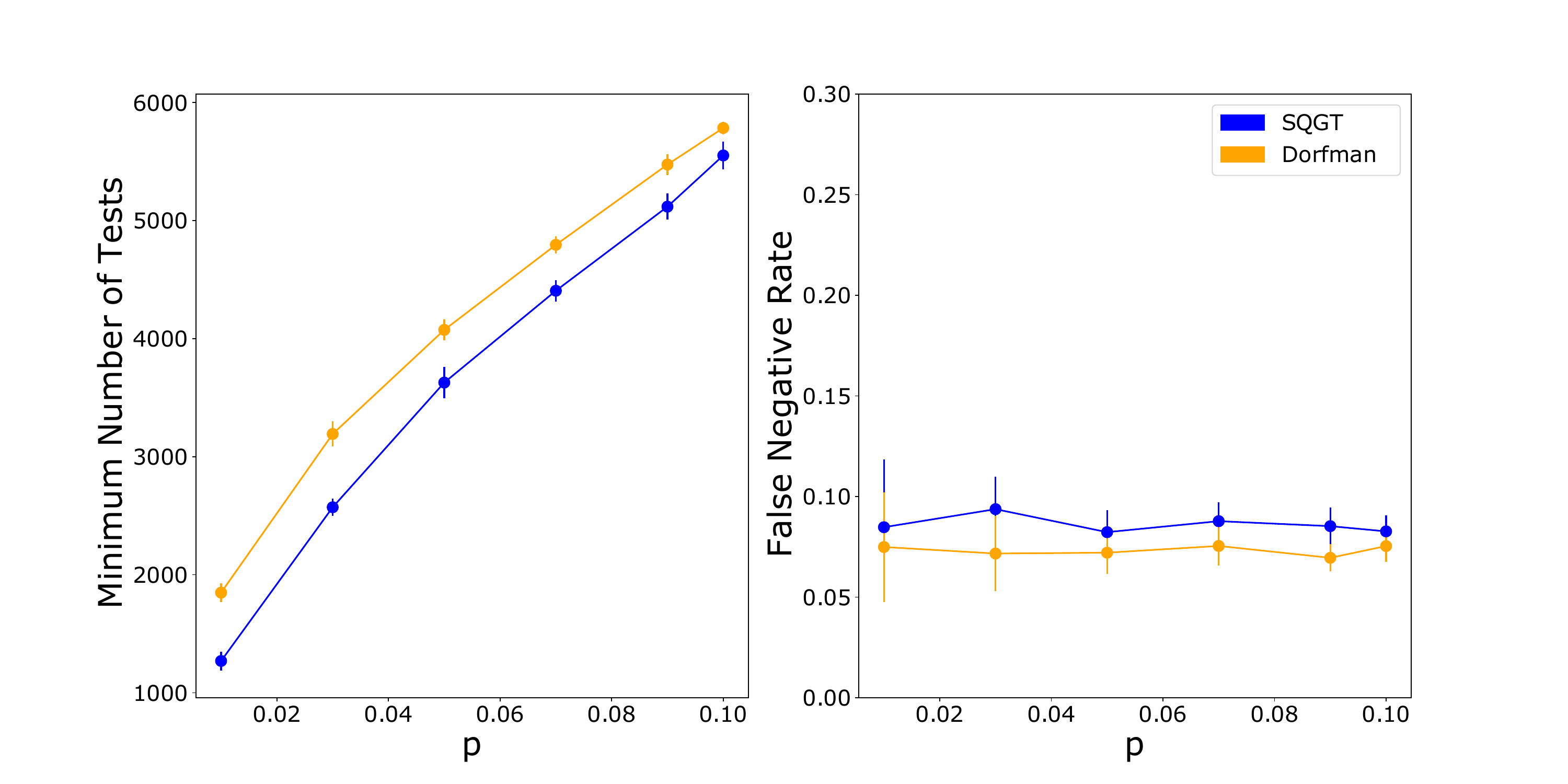}
    \caption{The optimal number of tests used in Dorfman's GT (orange) and SQGT (blue) versus the infection rate, $p$ (left panel), and the corresponding FNRs (right panel). Optimal refers to the smallest number of tests possible under all possible choices of group sizes $g$.} 
    \label{fig:sweep}
\end{figure}

Finally, we examined how the number of tests required for the optimal group size varies over a wider range of infection rates, as shown in Figure~\ref{fig:sweep}, alongside the corresponding FNRs. The figure shows that as the infection rate increases, the number of tests required for both GT schemes increases and the advantage of GT over individual testing decreases. This is a property that has been already established in the past for Dorfman's scheme~\cite{dorfman1943detection}. In addition, the figure shows that SQGT for PCR screening always saves more tests than Dorfman's scheme with only a small increase in FNR (within the margin of error of Dorfman's FNR).


\section*{Discussion}
We introduced the concept of Semi-Quantitative Group Testing (SQGT) as an extension of traditional Group Testing (GT) methods, with a specific focus on qPCR-based pathogen screening. GT methods, in their classical form, are based on binary test outcomes (positive or negative) and are effective for identifying infected individuals in a cost-efficient manner. However, they fail to utilize the full quantitative information provided by qPCR assays, which can lead to suboptimal performance in scenarios with widely varying pathogen concentrations.

SQGT addresses this limitation by interpreting test results as estimates of the number of infected individuals in each group. The proposed SQGT scheme utilizes two quantization thresholds to categorize qPCR results into different risk categories, allowing for a more refined analysis of the infection status within each group. By employing random permutations and two-stage testing, SQGT can reduce the number of tests needed while still maintaining a high level of test accuracy.

The study also addressed the issue of dilution effects in GT protocols, which can lead to false negatives in qPCR-based testing. To mitigate this problem, we incorporated group size-dependent thresholds in the SQGT framework, adjusting for the dilution effect and improving the overall accuracy of the test results.

Through extensive simulations and analysis using real-world qPCR data from SARS-CoV-2 testing, we demonstrated that SQGT outperforms traditional GT schemes (such as Dorfman's GT) in terms of test efficiency while maintaining a comparable or slightly higher FNR. For example, for a population infection rate of $p=0.02$, our conceptually simple SQGT method uses $24$\% fewer tests than the binary outcome Dorfman's GT method, while maintaining a negligible FNR compared to qPCR noise. In conclusion, SQGT provides substantial reductions in the number of tests required for pathogen screening, making it a promising approach for large-scale population testing, especially during pandemics or outbreaks. 

It is important to note that the proposed SQGT scheme is tailored specifically for qPCR testing and it involves two stages of testing, as originally suggested by Dorfman's scheme. The two stages are crucial for adaptive screening which informs the tests in the second stage based on the results in the first stage. Nonadaptive testing scheme, on the other hand, would result in potentially smaller delays of the test results but would require significantly more tests. They are also often too complicated to implement in practice as they require combinatorial sample mixing and decoding.  

Additionally, our studies were performed under two assumptions, error-free qPCR assays, and qPCR assays with a sigmoidal model of false negatives as a function of $Ct$ values. The incorporation of qPCR assay noise into the simulations led to a slight increase in FNRs, highlighting the need for careful consideration of assay accuracy for a broader range of practical pathogen detection schemes.

For other pathogens and datasets, our SQGT scheme can be modified as needed by combining adaptive and nonadaptive test schemes, including more than two thresholds, and integrating a specialized technique for identifying ``heavy hitters" (i.e., individuals with very high viral loads). These approaches are mathematically analyzed in the Supplement Section~\ref{supp:variableload}.



\section*{Methods}
\subsection*{Data}\label{sec:data}
The real-world COVID-19 GT data~\cite{barak2021lessons} used in this paper contains $133,816$ samples collected between April and September 2020 in Israel and tested experimentally via Dorfman's pooling. The original data contains the following information for each individual sample:
\begin{itemize}
    \item Sample id: A unique id for tracking the sample;
    \item Month: Information about the month when the sample is collected;
    \item Group id: An id indicating which group an individual sample belongs to in the test scheme. Samples within the same group share the same group id, and the test groups are of size $5$ and $8$;
    \item Result: Final test result for a sample (positive/negative);
    \item Sample viral $Ct$: $Ct$ value of an individual test. Note that this value is not available when the group test involving the sample is negative;
    \item Group viral $Ct$: $Ct$ value of the group to which the individual sample belongs to;
    \item Sample human $Ct$: $Ct$ value of an individual test for amplifying the human ERV-3~\cite{yuan2001quantification} gene. This $Ct$ value lying below a predetermined threshold serves as an internal control for whether a test was successful or not;
    \item Group human $Ct$: $Ct$ value of the group test used for amplifying the human ERV-3 gene.
\end{itemize}

As pointed out in the Results Section, there are some experimental inconsistencies between the results of the group tests and the individual tests. Specifically, in $70$ out of $1,887$ positive tests, the results of the group tests were positive while all individuals within the groups tested negative. These results can be explained as false positives for the group test, or as false negatives for the individual tests. We used this information to model the FNR of the dataset as described in our Results Section. Note that for simplicity we assume that there is only one positive individual sample within the group when a false negative result is recorded, as this is the most probable scenario. We hence use (Group test $Ct-M\log_{10}(g)$) as the estimated $Ct$ value for the individual test in the presence of a false negative, where $g$ as before denotes the group size, while $M\log_{10}(g)=2.895$. Our fitted model shown in Figure~\ref{fig:fnr} (a) is obtained through the MATLAB \texttt{fit} function.

\subsection*{Modelling COVID-19 group testing schemes}
\subsubsection*{Modelling PCR tests}
When modelling an individual test, individual $i$ with a viral load $v_i$ will have
\begin{linenomath}
\begin{align}\label{eq:lineari}
Ct_{i} = -M \log_{10}(v_i) + B.
\end{align}
\end{linenomath}
The values for $M$ and $B$ are set based on a previously established calibration curve \cite{jones2021estimating}. Then given a threshold $Ct_{I}$, an individual $i$ is considered positive for the virus if $Ct_{i}<\tau_{\textit{In}}$. In our simulations we use $\tau_{\textit{In}} = 36$. 

To model a pooled test, the viral loads of individuals in a group are averaged and plugged into Equation \ref{eq:linear} to determine the $Ct$ for the group. That is, for group $j$ with individuals $\{1,2,...,g\}$
\begin{linenomath}
\begin{align}\label{eq:linear}
Ct_{j} = -M \log_{10}(\frac{1}{g}\sum_{i=1}^{g} v_{ji}) + B.
\end{align}
\end{linenomath}
These group $Ct$s can then be used for different GT schemes as described in the Algorithms and Results section.
\subsection*{Including PCR noise into models}
Since PCR tests are not error-free, we also include some noise into the tests based on the FNR function
\begin{linenomath}
\begin{align}
FNR(Ct)=\left[1+\exp\left(\frac{a-Ct}{b}\right)\right]^{-1},
\end{align}
\end{linenomath}
where $b$ is empirically determined to be $2.145$ as discussed in the Algorithms and Results section and $a$ is the threshold used for the PCR test. To include this noise into our PCR simulations, we use the following procedure:

\begin{algorithmic}
\If{test is individual} 
    \State $Ct \gets -M \log_{10}(v_i) + B$
\Else
    \State $Ct \gets -M \log_{10}(\frac{1}{g}\sum_{i=1}^{g} v_{ji}) + B.$
    \EndIf
\State result $\gets$ Scheme($Ct$)
\If{truth == positive}
    \If{Bernoulli$(p=FNR(Ct))$}
    \State result $\gets$ negative
    \EndIf
\EndIf
\end{algorithmic}

First, the $Ct$ value of a test is calculated using Equation \ref{eq:lineari} or \ref{eq:linear}. If the ground truth of the test is that it is positive, it is converted into a negative (no infected individuals) with probability $FNR(Ct)$. Otherwise, the result of the test is left as determined by the testing scheme.


\section*{Acknowledgments}
We thank Professor Nigel Goldenfeld and Dr. Rayan Gabrys for useful discussions. The work was supported by NSF grants 2107344 and 2107345.

\section*{Author Contributions}
Conceptualization, S.M., and O.M.; Investigation, A.N., C.P., and V.R.; Formal Analysis, A.N., C.P., V.R., M.C., J.R., S.M., and O.M.; Writing – Original Draft, A.N., C.P., V.R., M.C., J.R., S.M., and O.M.; Writing – Review \& Editing, A.N., C.P., V.R., S.M., and O.M.;


\bibliography{main}

\begin{thebibliography}{10}

\bibitem{aeron2010information}
S.~Aeron, V.~Saligrama, and M.~Zhao.
\newblock Information theoretic bounds for compressed sensing.
\newblock {\em IEEE Transactions on Information Theory}, 56(10):5111--5130,
  2010.

\bibitem{alberts2002introduction}
B.~Alberts, A.~Johnson, J.~Lewis, M.~Raff, K.~Roberts, and P.~Walter.
\newblock Introduction to pathogens.
\newblock In {\em Molecular Biology of the Cell. 4th edition}. Garland Science,
  2002.

\bibitem{alcoba2021increasing}
J.~Alcoba-Florez, H.~Gil-Campesino, D.~G.-M. de~Artola, O.~D{\'\i}ez-Gil,
  A.~Valenzuela-Fern{\'a}ndez, R.~Gonz{\'a}lez-Montelongo, L.~Ciuffreda, and
  C.~Flores.
\newblock Increasing sars-cov-2 rt-qpcr testing capacity by sample pooling.
\newblock {\em International Journal of Infectious Diseases}, 103:19--22, 2021.

\bibitem{Ald18}
M.~{Aldridge}.
\newblock Individual testing is optimal for nonadaptive group testing in the
  linear regime.
\newblock {\em IEEE Transactions on Information Theory}, 65(4):2058--2061,
  2019.

\bibitem{baker2022infectious}
R.~E. Baker, A.~S. Mahmud, I.~F. Miller, M.~Rajeev, F.~Rasambainarivo, B.~L.
  Rice, S.~Takahashi, A.~J. Tatem, C.~E. Wagner, L.-F. Wang, et~al.
\newblock Infectious disease in an era of global change.
\newblock {\em Nature Reviews Microbiology}, 20(4):193--205, 2022.

\bibitem{barak2021lessons}
N.~Barak, R.~Ben-Ami, T.~Sido, A.~Perri, A.~Shtoyer, M.~Rivkin, T.~Licht,
  A.~Peretz, J.~Magenheim, I.~Fogel, et~al.
\newblock Lessons from applied large-scale pooling of 133,816 sars-cov-2 rt-pcr
  tests.
\newblock {\em Science Translational Medicine}, 13(589):eabf2823, 2021.

\bibitem{berger1998bounds}
T.~Berger and J.~W. Mandell.
\newblock Bounds on the efficiency of two-stage group testing.
\newblock In {\em Codes, Curves, and Signals: Common Threads in
  Communications}, pages 213--232. Springer, 1998.

\bibitem{brault2021group}
V.~Brault, B.~Mallein, and J.-F. Rupprecht.
\newblock Group testing as a strategy for covid-19 epidemiological monitoring
  and community surveillance.
\newblock {\em PLoS computational biology}, 17(3):e1008726, 2021.

\bibitem{BK20}
A.~Z. {Broder} and R.~{Kumar}.
\newblock A note on double pooling tests.
\newblock {\em arXiv e-prints}, Apr. 2020.

\bibitem{candes2006stable}
E.~J. Candes, J.~K. Romberg, and T.~Tao.
\newblock Stable signal recovery from incomplete and inaccurate measurements.
\newblock {\em Communications on Pure and Applied Mathematics: A Journal Issued
  by the Courant Institute of Mathematical Sciences}, 59(8):1207--1223, 2006.

\bibitem{cheraghchi2021semiquantitative}
M.~Cheraghchi, R.~Gabrys, and O.~Milenkovic.
\newblock Semiquantitative group testing in at most two rounds.
\newblock In {\em 2021 IEEE International Symposium on Information Theory
  (ISIT)}, pages 1973--1978. IEEE, 2021.

\bibitem{dai2009subspace}
W.~Dai and O.~Milenkovic.
\newblock Subspace pursuit for compressive sensing signal reconstruction.
\newblock {\em IEEE transactions on Information Theory}, 55(5):2230--2249,
  2009.

\bibitem{de2020sample}
A.~de~Salazar, A.~Aguilera, R.~Trastoy, A.~Fuentes, J.~C. Alados, M.~Causse,
  J.~C. Gal{\'a}n, A.~Moreno, M.~Trigo, M.~P{\'e}rez-Ruiz, et~al.
\newblock Sample pooling for sars-cov-2 rt-pcr screening.
\newblock {\em Clinical Microbiology and Infection}, 26(12):1687--e1, 2020.

\bibitem{donoho2006compressed}
D.~L. Donoho.
\newblock Compressed sensing.
\newblock {\em IEEE Transactions on information theory}, 52(4):1289--1306,
  2006.

\bibitem{dorfman1943detection}
R.~Dorfman.
\newblock The detection of defective members of large populations.
\newblock {\em The Annals of mathematical statistics}, 14(4):436--440, 1943.

\bibitem{Dor43}
R.~Dorfman.
\newblock The detection of defective members of large populations.
\newblock {\em The Annals of Mathematical Statistics}, 14(4):436--440, 1943.

\bibitem{du2000combinatorial}
D.~Du, F.~K. Hwang, and F.~Hwang.
\newblock {\em Combinatorial group testing and its applications}, volume~12.
\newblock World Scientific, 2000.

\bibitem{d1984generalized}
A.~Dyachkov and V.~Rykov.
\newblock Generalized superimposed codes and their application to random
  multiple access.
\newblock In {\em Proc. 6th Int. Symp. Inf. Theory}, volume~1, pages 62--64,
  1984.

\bibitem{eberhardt2020multi}
J.~N. Eberhardt, N.~P. Breuckmann, and C.~S. Eberhardt.
\newblock Multi-stage group testing improves efficiency of large-scale covid-19
  screening.
\newblock {\em Journal of Clinical Virology}, 128:104382, 2020.

\bibitem{emad2014semiquantitative}
A.~Emad and O.~Milenkovic.
\newblock Semiquantitative group testing.
\newblock {\em IEEE Transactions on Information Theory}, 60(8):4614--4636,
  2014.

\bibitem{emad2016code}
A.~Emad and O.~Milenkovic.
\newblock Code construction and decoding algorithms for semi-quantitative group
  testing with nonuniform thresholds.
\newblock {\em IEEE Transactions on Information Theory}, 62(4):1674--1687,
  2016.

\bibitem{fajnzylber2020sars}
J.~Fajnzylber, J.~Regan, K.~Coxen, H.~Corry, C.~Wong, A.~Rosenthal, D.~Worrall,
  F.~Giguel, A.~Piechocka-Trocha, C.~Atyeo, et~al.
\newblock Sars-cov-2 viral load is associated with increased disease severity
  and mortality.
\newblock {\em Nature communications}, 11(1):1--9, 2020.

\bibitem{fortuin1971correlation}
C.~M. Fortuin, P.~W. Kasteleyn, and J.~Ginibre.
\newblock Correlation inequalities on some partially ordered sets.
\newblock {\em Communications in Mathematical Physics}, 22(2):89--103, 1971.

\bibitem{fraser2007variation}
C.~Fraser, T.~D. Hollingsworth, R.~Chapman, F.~de~Wolf, and W.~P. Hanage.
\newblock Variation in hiv-1 set-point viral load: epidemiological analysis and
  an evolutionary hypothesis.
\newblock {\em Proceedings of the National Academy of Sciences},
  104(44):17441--17446, 2007.

\bibitem{ghosh2021compressed}
S.~Ghosh, R.~Agarwal, M.~A. Rehan, S.~Pathak, P.~Agarwal, Y.~Gupta, S.~Consul,
  N.~Gupta, R.~Goenka, A.~Rajwade, et~al.
\newblock A compressed sensing approach to pooled rt-pcr testing for covid-19
  detection.
\newblock {\em IEEE Open Journal of Signal Processing}, 2:248--264, 2021.

\bibitem{hogan2020sample}
C.~A. Hogan, M.~K. Sahoo, and B.~A. Pinsky.
\newblock Sample pooling as a strategy to detect community transmission of
  sars-cov-2.
\newblock {\em Jama}, 323(19):1967--1969, 2020.

\bibitem{hu2020factors}
X.~Hu, Y.~Xing, J.~Jia, W.~Ni, J.~Liang, D.~Zhao, X.~Song, R.~Gao, and
  F.~Jiang.
\newblock Factors associated with negative conversion of viral rna in patients
  hospitalized with covid-19.
\newblock {\em Science of the Total Environment}, 728:138812, 2020.

\bibitem{hwang1976group}
F.~K. Hwang.
\newblock Group testing with a dilution effect.
\newblock {\em Biometrika}, 63(3):671--680, 1976.

\bibitem{indyk2010efficiently}
P.~Indyk, H.~Q. Ngo, and A.~Rudra.
\newblock Efficiently decodable non-adaptive group testing.
\newblock In {\em Proceedings of the twenty-first annual ACM-SIAM symposium on
  Discrete Algorithms}, pages 1126--1142. SIAM, 2010.

\bibitem{jones2021estimating}
T.~C. Jones, G.~Biele, B.~M{\"u}hlemann, T.~Veith, J.~Schneider,
  J.~Beheim-Schwarzbach, T.~Bleicker, J.~Tesch, M.~L. Schmidt, L.~E. Sander,
  et~al.
\newblock Estimating infectiousness throughout sars-cov-2 infection course.
\newblock {\em Science}, 373(6551):eabi5273, 2021.

\bibitem{lin2020group}
Y.~Lin, Y.~Ren, J.~Wan, M.~Cashore, J.~Wan, Y.~Zhang, P.~Frazier, and E.~Zhou.
\newblock Group testing enables asymptomatic screening for covid-19 mitigation:
  Feasibility and optimal pool size selection with dilution effects.
\newblock {\em arXiv preprint arXiv:2008.06642}, 2020.

\bibitem{lindstrom1975determining}
B.~Lindstrom.
\newblock Determining subsets by unramified experiments.
\newblock {\em A Survey of Statistical Design and Linear Models}, 1975.

\bibitem{TAKAHASHI2022e01496}
H.~Takahashi, N.~Ichinose, and Y.~Okada.
\newblock False-negative rate of sars-cov-2 rt-pcr tests and its relationship
  to test timing and illness severity: A case series.
\newblock {\em IDCases}, 28:e01496, 2022.

\bibitem{wolf1985born}
J.~Wolf.
\newblock Born again group testing: Multiaccess communications.
\newblock {\em IEEE Transactions on Information Theory}, 31(2):185--191, 1985.

\bibitem{woloshin2020false}
S.~Woloshin, N.~Patel, and A.~S. Kesselheim.
\newblock False negative tests for sars-cov-2 infection—challenges and
  implications.
\newblock {\em New England Journal of Medicine}, 383(6):e38, 2020.

\bibitem{yelin2020evaluation}
I.~Yelin, N.~Aharony, E.~S. Tamar, A.~Argoetti, E.~Messer, D.~Berenbaum,
  E.~Shafran, A.~Kuzli, N.~Gandali, O.~Shkedi, et~al.
\newblock Evaluation of covid-19 rt-qpcr test in multi sample pools.
\newblock {\em Clinical Infectious Diseases}, 71(16):2073--2078, 2020.

\bibitem{yuan2001quantification}
C.~C. Yuan, W.~Miley, and D.~Waters.
\newblock A quantification of human cells using an erv-3 real time pcr assay.
\newblock {\em Journal of virological methods}, 91(2):109--117, 2001.

\bibitem{zheng2020viral}
S.~Zheng, J.~Fan, F.~Yu, B.~Feng, B.~Lou, Q.~Zou, G.~Xie, S.~Lin, R.~Wang,
  X.~Yang, et~al.
\newblock Viral load dynamics and disease severity in patients infected with
  sars-cov-2 in zhejiang province, china, january-march 2020: retrospective
  cohort study.
\newblock {\em bmj}, 369, 2020.

\end{thebibliography}

\bibliographystyle{abbrv}

\pagebreak
\setcounter{figure}{0}
\renewcommand{\figurename}{Supplement Figure}
{\Large
}
\section{Supplementary Information}
\subsection{Supplementary Figures}
\begin{figure}[ht!]
     \centering
     \includegraphics[width=0.85\textwidth]{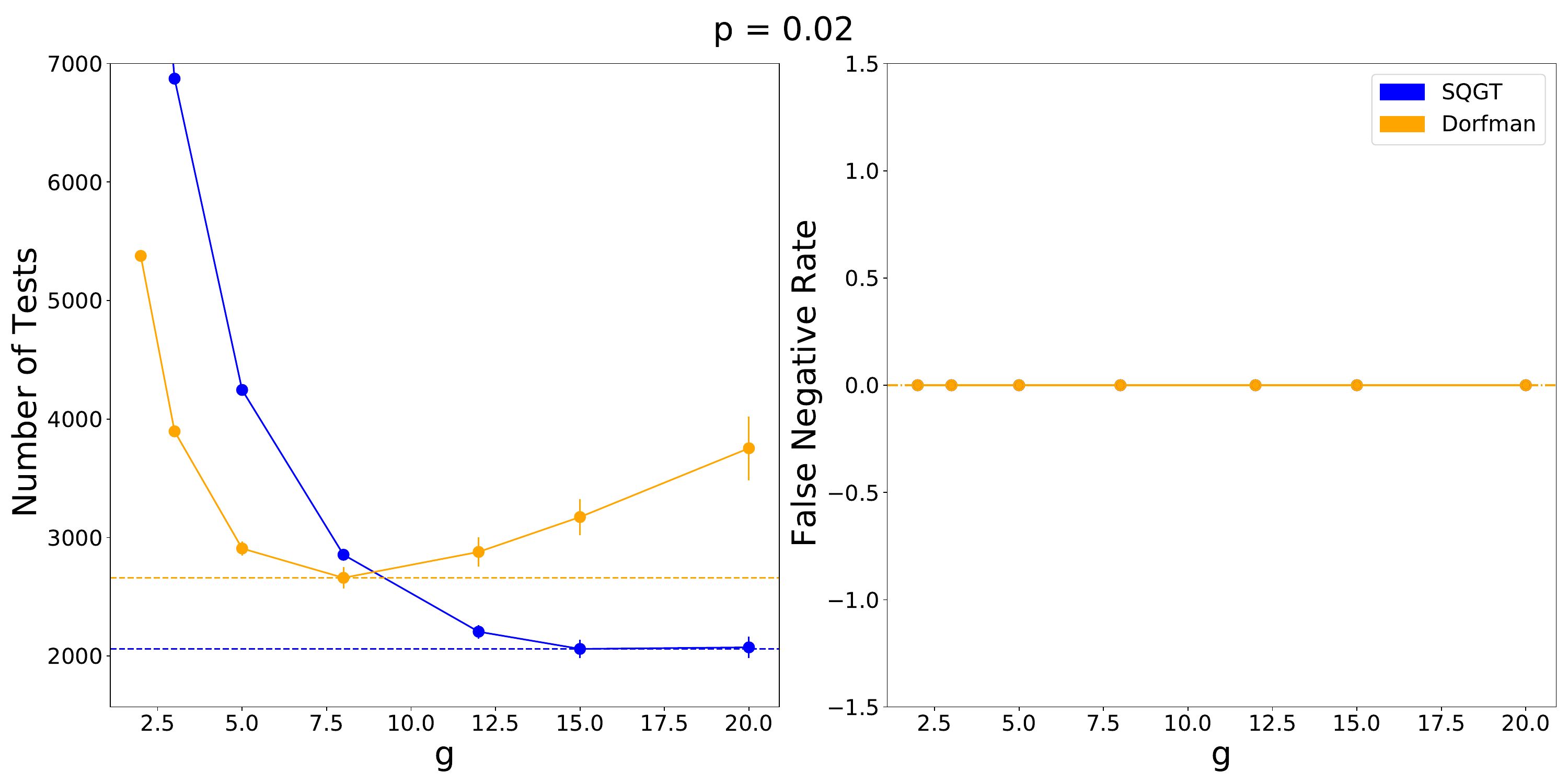}\\
     \includegraphics[width=0.85\textwidth]{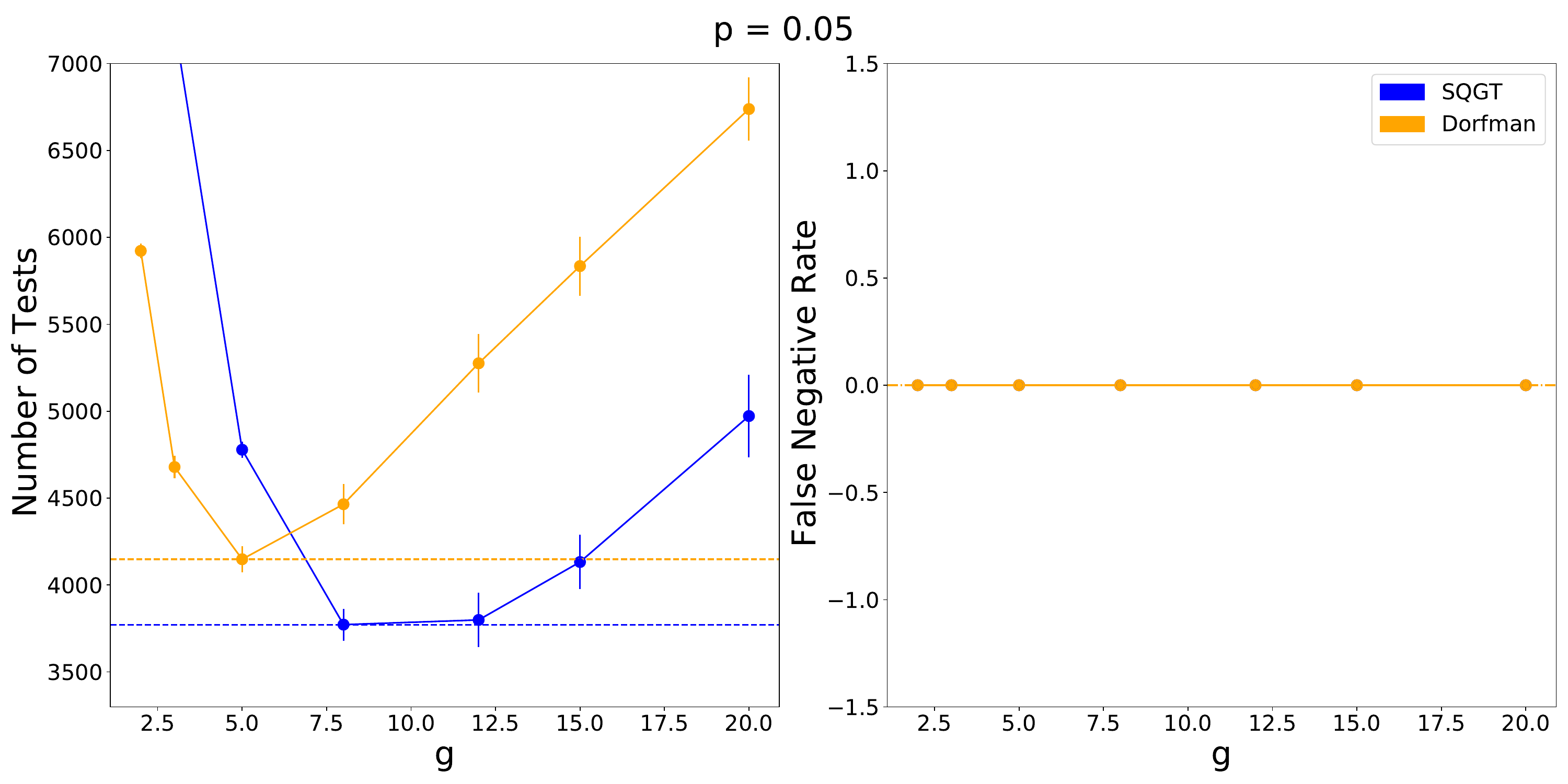}\\
     \includegraphics[width=0.85\textwidth]{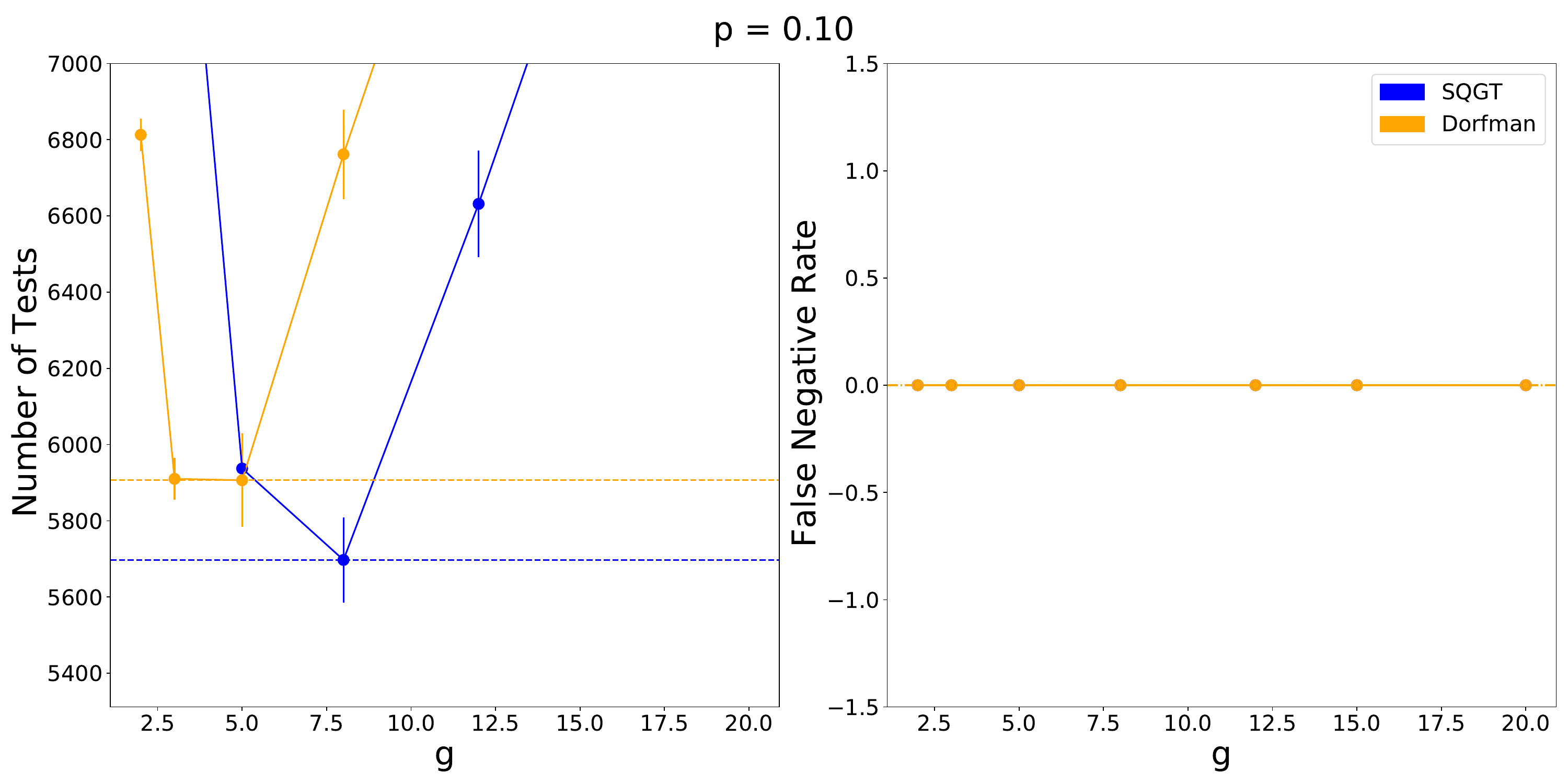}
     \caption{The number of tests used and the false negative rates using error-free PCR of SQGT (blue), Dorfman's GT (orange) and individual testing (red) for infection rates $p\in \{0.02,0.05,0.1\}.$ The dashed line marks indicate the number of tests and false negative rates for the optimal group size (where the number of tests is minimized) for each scheme.}
     \label{fig:error-free}
 \end{figure}

    \subsection{Analysis of GT models}\label{supp:modelGT}

    Group testing (GT) schemes based on binary test outcomes, positive $(1)$ or negative $(0),$ and using single~\cite{dorfman1943detection} or multiple rounds~\cite{berger1998bounds} of pooling can be mathematically analyzed in a straightforward manner. Once again, we consider a population of $n$ individuals and denote the fraction of infected individuals by $p$. We also assume that the test results are accurate, i.e., the FNR and FPR are identically zero.

    For the single-pooling Dorfman's GT, we randomly split the $n$ individuals into groups of size $g$. Let $E[T/n]$ be the expected number of tests performed per individual under this scheme. In the first stage, we use $n/g$ tests, resulting in $1/g$ tests per individual. In the second stage, individuals in groups with a positive outcome are tested individually, and the rest are declared as ``not infected". A group test outcome is positive if at least one infected individual is among the $g$ individuals in the group, which occurs with a probability of $1-(1-p)^g.$ Therefore the expected number of tests per individual in the second stage is $$0*(1-p)^g + 1*(1-(1-p)^g) = 1-(1-p)^g,$$
    and the expected number of tests per individual for the entire scheme is
    $$E[T/n] = \frac{1}{g} + 1-(1-p)^g.$$

    We can perform a similar analysis for a double-pooling scheme. We group the individuals into groups of size $g$ such that each individual contributes to two different groups. This can either be achieved via two random permutations of the list of individuals, similar to SQGT, or via a 2D array of wells with appropriately chosen dimensions, as described in~\cite{berger1998bounds}. Therefore, for each individual, we have a pair of test outcomes. If both the test outcomes are positive $(1),$ we test individually otherwise we declare the individual as not infected.

    In the first stage, we use $2n/g$ tests, resulting in $2/g$ tests per individual. In the second stage, an individual is tested if both group tests are positive. We consider two scenarios: the individual is infected with probability $p$, or the individual is not infected with probability $1-p$. If the individual is infected, both groups they contributed to will test positive, leading to testing in the second stage. If the individual is not infected, the group test outcome is positive if and only if at least one other person in the group is infected. For the individual to get tested in the second stage, this condition must hold for both groups they are part of in the first stage. This allows us to calculate the expected number of tests in the second stage as
    $$1*(p + (1-p)(1-(1-p)^{g-1})^2).$$
    Therefore, the expected number of tests per individual is
    $$E[T/n] = 2/g + p + (1-p)(1-(1-p)^{g-1})^2.$$

    This approach can be extended to a multi-pooling strategy, with each individual contributing to more than two group tests. However, such schemes have diminishing returns in the number of tests saved due to substantial overlaps in the groups~\cite{berger1998bounds}.

    Our Semi-Quantitative Group Testing (SQGT) scheme extends the double-pooling strategy by utilizing semi-quantitative information from a qPCR test. This scheme improves performance by avoiding the direct individual testing of low and medium-risk individuals. Under the assumption that the FNR of qPCR testing is zero, our relaxed SQGT scheme (see Results) yields the same results as double-pooling.
    
    \subsection{Probabilistic SQGT with variable viral load}\label{supp:variableload} We analyze how the semi-quantitative scheme 
    performs when infected individuals may have either low or high viral loads. This is relevant to account for heavy hitters, individuals with substantially higher viral loads which can mask infected individuals with low viral loads. To this end, we consider a simplified model where each individual is independently infected and presents a low viral load at the time of testing with probability $p_{i1}$, or is infected and presents a high viral load at the time of testing with probability $p_{i2}.$ In particular, each individual is infected (regardless of their viral load) with total infection probability $p=p_{i1}+p_{i2}<1$.

    Individuals with high viral loads are problematic because, based on the semi-quantitative output of qPCR, groups featuring \emph{one} such individual may be mistaken for groups with \emph{several} infected individuals with low-to-intermediate viral loads.\footnote{This is not problematic for \emph{binary} GT, where the test outcomes do not distinguish between one or several infected individuals in the group.}
    This phenomenon naturally leads us to consider the following modified version of testing: A test applied to a group of individuals has outcome $0$ if there are no infected individuals in the group, outcome $1$ if there exists \emph{exactly} one infected individual with \emph{low} viral load, and $2$ if either there exists more than one infected individual with low viral load or at least one infected individual with \emph{high} viral load. Therefore, as expected, individuals with high viral load obfuscate the test outcomes.

    We assume that the population contains $n$ individuals, each of which is independently positive with some probability $p=p_{i1}+p_{i2}<1$, as explained.  In the first stage, we divide the $n$ individuals into groups of size $g.$ The groups are denoted by $\gamma_1,\gamma_2,\ldots, \gamma_{n/g}.$ In the second stage, we proceed as follows:\\
    \begin{itemize}
        \item If a pool $\gamma_i$ tests $0$, we declare all individuals in $\gamma_i$ as negative. \\
        \item If a pool $\gamma_i$ tests $1$, we apply a nearly-optimal zero-error nonadaptive GT scheme to detect the infected individual.\\
        \item If a pool $\gamma_i$ tests $2$, we test all individuals in $\gamma_i$ separately.\\
    \end{itemize}
    We can compute the expected number of tests per individual of the testing scheme, $E[T/n]$, as a function of the probability of infection $p$ and the first-stage pool size $g$ as follows. 
    First, we observe that for the scheme outlined above, we are guaranteed to have \emph{exactly} $1$ infected individual in any pool that tested $1$. We also know that zero-error nonadaptive GT schemes to detect $\tau$ infected individuals in a group of size $g$ can be designed with $m(g,\tau)=c\cdot \tau^2\log(g/\tau)$ tests for some constant $c>0$.
    As a corollary, we know that for detecting one infected individual, $m(g,1)=\lceil \log g\rceil$ tests are needed. 
    This can be achieved by using a Hamming code parity-check matrix. This gives us the expected number of tests as
    \begin{equation}\label{eq:expsqgt}
	\E[T/n]=\frac{1}{g}+p_1\cdot \lceil \log g\rceil+p_2,
    \end{equation}
    where $p_1$ and $p_2$ denote the probability that a given pool tests $1$ and~$2$, respectively.

    The probability that a group of size $g$ contains exactly one infected individual with low viral load and zero individuals with high viral load (leading to test outcome $1$) is
    \begin{equation*}
    	p_1 = g\cdot p_{i1} \cdot (1-p_{i1}-p_{i2})^{g-1}=g \cdot p_{i1} \cdot (1-p)^{g-1},
    \end{equation*}
    while the probability that the group contains either more than one infected individual with low viral load or at least one individual with high viral load (leading to test outcome $2$) is
    \begin{equation*}
    	p_2 = 1-g \cdot p_{i1} \cdot (1-p_{i1}-p_{i2})^{g-1}-(1-p_{i1}-p_{i2})^g=1-g \cdot p_{i1} \cdot (1-p)^{g-1}-(1-p)^g.
    \end{equation*}
	
    Combining these observations, we conclude that the expected number of tests per individual as a function of $p_{i1}$ and $p_{i2}$ is given by
    \begin{equation}\label{eq:expSQGTviral}
    	\frac{1}{g}+g \cdot p_{i1} \cdot (1-p)^{g-1} \cdot \lceil\log g\rceil +1-g \cdot p_{i1} \cdot (1-p)^{g-1}-(1-p)^g,
    \end{equation}
    where $p=p_{i1}+p_{i2}$.
    
    For fixed $p_{i1}$ and $p_{i2}$, it is easy to numerically minimize the expression above as a function of $g$ to find the optimal group size for the scheme under consideration. On the other hand, the expected number of tests per individual for the basic Dorfman's GT~\cite{Dor43} is
	\begin{equation}\label{eq:expsp}
	\frac{1}{g}+1-(1-p)^g,
	\end{equation}
	and the expected number of tests per individual for a double-pooling scheme with binary tests~\cite{BK20, berger1998bounds} is
	\begin{equation}\label{eq:expdp}
	\frac{2}{g}+p+(1-p)(1-(1-p)^{g-1})^2.
	\end{equation}	
    
\begin{figure}[h]
    	\centering
    	\includegraphics[width=0.9\textwidth]{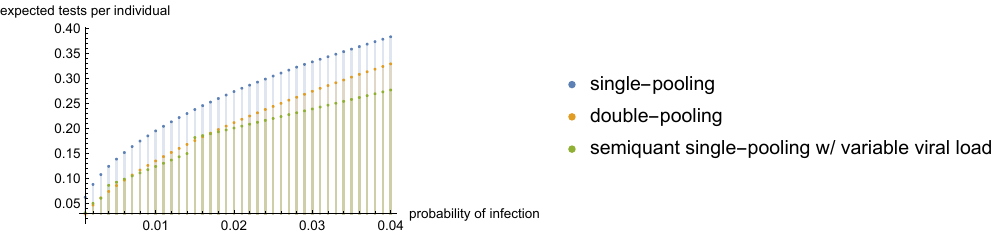}
    	\caption{Comparison between the expected number of tests per individual required by Dorfman's single pooling scheme~\cite{Dor43}, the double pooling scheme~\cite{BK20, berger1998bounds}, and our semi-quantitative single pooling scheme as a function of total infection probability $p$ with $p_{i1}=0.84p,\,p_{i2}=0.16p$.}
    	\label{fig:comp-sp-dp-sqsp-viral16}
    \end{figure}

\begin{figure}[h]
    	\centering
    	\includegraphics[width=0.9\textwidth]{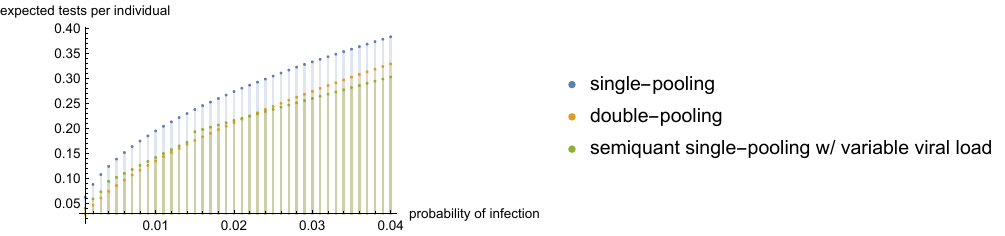}
    	\caption{Comparison between the expected number of tests per individual required by Dorfman's single pooling scheme~\cite{Dor43}, the double pooling scheme~\cite{BK20, berger1998bounds}, and our semi-quantitative single pooling scheme as a function of total infection probability $p$ with $p_{i1}=2p/3$ and $p_{i2}=p/3$.}
    	\label{fig:comp-sp-dp-sqsp-viralthird}
    \end{figure}

    Figures~\ref{fig:comp-sp-dp-sqsp-viral16} and~\ref{fig:comp-sp-dp-sqsp-viralthird} compare the expected number of tests per individual required by various schemes for different values of the total infection probability $p$ and the specific infection probabilities $p_{i1}$ and $p_{i2}$. Clearly, double pooling outperforms Dorfman's single pooling GT strategy, while semi-quantitative testing with single pooling outperforms both single and double pooling in the expected number of tests. SQGT combines the ideas of double pooling and semi-quantitative information from tests to obtain further savings in test results. We do not have a closed-form expression for SQGT due to the complexity of the scheme. However, as reported in the Results, double pooling SQGT provides substantial savings over Dorfman's GT (single pooling) while maintaining low FNR for real-world GT data. 


    \subsection{Lower bounds for nonadaptive probabilistic SQGT}\label{supp:nonadaptiveSQGT} 
    The main text discusses how to replace the second stage of SQGT with nonadaptive GT. Designing nonadaptive GT schemes reduces to identifying test-matrices that satisfy disjointness or separability properties~\cite{du2000combinatorial}. One such scheme involves sampling a random binary matrix such that every entry is an i.i.d. Bernoulli($q$) random variable with $0<q<1.$ Each row of this matrix defines a pool for group tests. Given sufficiently many rows, this matrix represents a zero-error nonadaptive GT scheme with high probability.
     
    We, therefore, focus on deriving a theoretical result that establishes lower bounds for nonadaptive probabilistic GT that may be used to assess the quality of our adaptive schemes. For this purpose, we adapt an argument by Aldridge~\cite{Ald18} for arbitrarily small error probability under a constant probability of infection.
	More precisely, we consider a setting where each test has $m+1$ outcomes for some $m \geq 1$: The outcome of a test is either $i$ if there are exactly $i$ infected individuals for $i<m$, and $\geq m$ otherwise. This corresponds to the setting introduced in~\cite{d1984generalized} which provides the most informative type of measurements one can expect from the SQGT framework using the amplification curve information. This model accounts for the saturation limit for each test, dictated by $m$, which is a phenomenon observable from the amplification curve. Moreover, as before we assume that each individual in the population of size $n$ is infected independently with some constant probability $p>0$.
	We show the following.
	\begin{thm}\label{thm:lb}
		For every $m$ and constant $p>0$ there exists a constant $\eps(m,p)>0$ such that, under the setting described above, nonadaptive testing requires at least $n/m$ tests to achieve error probability less than $\eps(m,p)$ in a population of size $n$.
	\end{thm}
	In contrast, for $m=2$, our two-stage scheme uses significantly fewer than $n/2$ tests provided $p$ is not very large.
	
	Proving Theorem~\ref{thm:lb} follows by a simple adaptation of an approach by Aldridge~\cite{Ald18}, who showed that individual testing is required in order to achieve arbitrarily small error in regular nonadaptive probabilistic GT (which corresponds to $m=1$).
	First, given any nonadaptive testing scheme, we may without loss of generality remove all tests with $m$ or fewer elements, along with all individuals who participate in those tests. This does not affect the lower bound. Then, we show that there are no nonadaptive testing schemes with an arbitrarily small error where every test includes at least $m+1$ individuals.
	Combining these two observations immediately yields Theorem~\ref{thm:lb}.
	
	For an individual $i$, let $x_i$ denote its infection status.
	Call an individual $i$ (regardless of its infection status) \emph{disguised} if every test $t$ in which it participates contains at least $m$ other individuals which are infected.
	If $i$ is disguised, then changing $x_i$ from $0$ to $1$, or vice-versa, does not change the outcome of the testing scheme.
	As a result, we can do no better than guess $x_i$, and we will be wrong with probability at least $\min(p,1-p)$.
	To finalize the argument, it suffices to show there is a disguised individual with constant probability.
	
	Let $D_i$ denote the event that individual $i$ is disguised, and let $D_{t,i}$ denote the event that individual $i$ is disguised in test $t$.
	Since the $D_{t,i}$ are increasing events\footnote{If $D_{t,i}$ holds and the set of infected individuals is expanded, then $D_{t,i}$ continues to hold under this expanded set.}, the Fortuin-Kasteleyn-Ginibre (FKG) inequality~\cite{fortuin1971correlation} implies that
	\begin{equation}
	\Pr[D_i]\geq \prod_{t:x_{t,i}=1}\Pr[D_{t,i}],
	\end{equation}
	where $x_{t,i}$ indicates whether individual $i$ participates in test $t$.
	Moreover, we have
	\begin{equation}
	\Pr[D_{t,i}]=\Pr[B(w_t-1,p)\geq r],
	\end{equation}
	where $w_t=\sum_{i=1}^n x_{t,i}$ is the weight of test $t$ and $B(w_t-1,p)$ denotes a binomial random variable with $w_t-1$ trials and success probability $p$.
	
        Let
	\begin{equation*}
	L_i=\log\left(\prod_{t:x_{t,i}=1}\Pr[D_{t,i}]\right)=\sum_{t:x_{t,i}=1}\log \Pr[D_{t,i}]=\sum_{t=1}^T x_{t,i}\log \Pr[D_{t,i}],
	\end{equation*}
	where $T$ denotes the total number of tests, which we assume satisfies $T/n<1$, and $\log$ denotes $\log_2.$
	Then, it suffices to show that there exists some $i^\star$ with $L_{i^\star}>c$ for some constant $c$ independent of $n$.
	Let $I$ be uniformly distributed over $\{1,2,\dots,n\}$, and let $\overline{L}=\E[L_I]$.
	We have
	\begin{align*}
	\overline{L}&=\frac{1}{n}\sum_{i=1}^n \sum_{t=1}^T x_{t,i}\log \Pr[D_{t,i}]\\
	&=\frac{1}{n}\sum_{t=1}^T w_t\log \Pr[D_{t,i}]\\
	&\geq \min_{t=1,\dots,T} w_t\log \Pr[B(w_t-1,p)\geq m]\\
	&\geq\min_{w\geq r+1} w\log \Pr[B(w-1,p)\geq r]=:L^\star,
	\end{align*}
	where the second equality follows from the fact that $\Pr[D_{t,i}]$ is the same for every $i$ such that $x_{t,i}=1$, and in the first inequality we use the assumption that $T/n<1$.
	It is immediate that there exists some $i^\star$ with $L_{i^\star}\geq \overline{L}$, which implies that $\Pr[D_{i^\star}]\geq 2^{L^\star}$.
	Therefore, the error probability of the testing scheme is at least $\eps(m,p)=\min(p,1-p)\cdot 2^{L^\star}$.
	Noting that $L^\star$ does not depend on $n$ and is bounded from below for any $m$ and $p$ concludes the proof (since $\lim_{w\to\infty} w\log \Pr[B(w-1,p)\geq m]=1$). 

    \subsection{Extension of Hwang's model~\cite{hwang1976group} to SQGT}\label{supp:hwangFPR}

    \textbf{Definition of $TPR_1(p,g)$.}
    We can define the conditional probability $TPR_1(p,g)$ following the same idea as in Hwang's paper as 
    \begin{align}
        TPR_1(p,g)&=\P(\text{test score is }1|\text{there is exactly $1$ positive subject in the group})\nonumber\\
        &=\frac{A(p,g)}{g\cdot p\cdot (1-p)^{g-1}},
    \end{align}
    where $A(p,g)$ is chosen such that $TPR_1(k)$ satisfies the following two limit conditions given the infection rate $p<0.5$:
    \begin{align}\label{eq:limits}
        TPR_1(p,1) = 1, TPR_1(p,\infty) = 0.
    \end{align}
    Specifically, $TPR_1(p,\infty) = 0$ holds since there will be only $1$ infection in this group of size infinity. Based on~(\ref{eq:limits}), one simple form of $A(p,g)$ can be $A(k)=p^g$, which implies 
    \begin{align}\label{eq:tpr1}
        TPR_1(p,g)=\frac{p^g}{g\cdot p\cdot (1-p)^{g-1}}.
    \end{align}
    When $g=1, TPR_1(p, 1)=1$; when $g\rightarrow\infty$, $TPR_1(p,g)=\frac{1}{g}(\frac{p}{1-p})^{g-1}$. Since we assume $\frac{p}{1-p}<1$, so $TPR_1(p,\infty)\rightarrow 0$.
    
    When taking the dilution effect into consideration, we introduce the coefficient $d$ as in~\cite{hwang1976group}. When $d=0$, there is no dilution effect, meaning that $TPR_1(p,g)=1$ for every choice of group size $g$; when $d=1$, the dilution is complete and the probability should be of the form~(\ref{eq:tpr1}). Therefore, the final expression for $TPR_1(p,g)$ with dilution effects would be
    \begin{align}\label{eq:tpr1_dilute}
        TPR_1(p,g,d)=\frac{p^{g^{d}}}{g^{d}\cdot p\cdot (1-p)^{g^{d}-1}}.
    \end{align}
    
    \noindent\textbf{Definition of $TPR_2(p,g)$.}
    We can define the conditional probability $TPR_2(p,g)$ as 
    \begin{align}
        TPR_2(p,g)&=\P(\text{test score is 1 or 2}|\text{there are at least $2$ positive subjects in the group})\nonumber\\
        &=\frac{B(k)}{1-(1-p)^g-g\cdot p\cdot (1-p)^{g-1}}.
    \end{align}
    The two limit conditions are
    \begin{align}\label{eq:limits2}
        TPR_2(p,2) = 1, TPR_2(p,\infty) = p.
    \end{align}
    It is worth pointing out that the limiting of $TPR_2(p,\infty)$ is very different from $TPR_1(p,\infty)$, since now there can be many infections in this huge group, so the limiting probability will be approximately the probability of sampling a subject from the population uniformly at random and the subject being positive. In this case, we can have 
    \begin{align}\label{eq:tpr2}
        TPR_2(p,g)=\frac{p^{1+2/g}}{1-(1-p)^g-g\cdot p\cdot (1-p)^{g-1}}.
    \end{align}
    It is easy to check that the limits~(\ref{eq:limits2}) hold for $TPR_2(p,g)$.
    
    When taking the dilution effect into consideration, we can again make use of the dilution coefficient $d$. The final expression for $TPR_2(p,g,d)$ would be
    \begin{align}\label{eq:tpr2_dilute}
        TPR_2(p,g,d)=\frac{p^{1+(g/2)^{-d}}}{1-(1-p)^{2(g/2)^{d}}-2(g/2)^{d}\cdot p\cdot (1-p)^{2(g/2)^{d}-1}}.
    \end{align}
    
    \noindent\textbf{Expected cost.}
    With $TPR_1(p,g,d)$ and $TPR_2(p,g,d)$, we can compute the expected cost of one round of SQGT test. The expected number of infections in a group, when there is only $1$ infection in the group, is always $1$. Meanwhile, the expected number of infections in a group, when there are at least $2$ infections in the group and the group size is $g$, is 
    \begin{align*}
        \E(\text{infections}|\text{\# of infections}\geq 2)&=\sum_{i=2}^g i\cdot \P(\text{\# of infections = }i|\text{\# of infections}\geq 2) \\
        &=\frac{\sum_{i=2}^g i\cdot \P(\text{\# of infections = }i)}{\P(\text{\# of infections}\geq 2)}.
    \end{align*}
    Note that $\sum_{i=0}^g i\cdot \P(\text{\# of infections = }i)=pg$ based on the binomial distribution, so $\sum_{i=2}^g i\cdot \P(\text{\# of infections = }i)=pg-pg(1-p)^{g-1}=pg(1-(1-p)^{g-1})$. We have
    \begin{align}
        \E(\text{infections}|\text{\# of infections}\geq 2)=\frac{pg(1-(1-p)^{g-1})}{1-(1-p)^g-pg(1-p)^{g-1}}.
    \end{align}
    So the total expected cost is
    \begin{align*}
        E_{\text{cost}}(g)=&\;\frac{n}{g}\cdot (\text{expected cost for each group of size } g)\\
        =&\;\frac{n}{g}\cdot \big\{1+pg(1-p)^{g-1}\left(TPR_1(p,g,d)\cdot g+(1-TPR_1(p,g,d))\cdot c\right)+\\
        &(1-(1-p)^g-pg(1-p)^{g-1})(TPR_2(p,g,d)\cdot g+\\
        &(1-TPR_2(p,g,d))\cdot c\cdot \E(\text{infections}|\text{\# of infections}\geq 2))\big\},
    \end{align*}
    where $c$ is the cost if we misidentify a positive group, and for simplicity set the lab cost for each test to $\$1$. By taking the derivative of $E_{\text{cost}}(g)$ to $0$ we can get the optimal group size $g$ in this case.

\newpage
\end{document}